\documentclass[12pt]{article}
\pdfoutput=1
\usepackage{graphicx}
\graphicspath{{Figures/}{../Figures/}}

\usepackage{amsmath}
\usepackage{graphicx,psfrag,epsf}
\usepackage{enumerate}
\usepackage{natbib}
\usepackage{url} 
\usepackage{hyperref}
\usepackage{graphicx} 
\usepackage{amsthm,amssymb}
\usepackage{bm}
\usepackage{hyperref}
\usepackage{dsfont}
\usepackage[caption=false]{subfig}
\usepackage{tabularx}
\usepackage{float}
\usepackage{algorithm}
\usepackage[noend]{algpseudocode}
\usepackage[section]{placeins}
\usepackage{multirow}
\usepackage{todonotes}

\begin{document}

	\def\spacingset#1{\renewcommand{\baselinestretch}%
		{#1}\small\normalsize} \spacingset{1}


		\title{ A purely data-driven framework for prediction, optimization, and control of networked processes: application to networked SIS epidemic model }
		\author{ Ali Tavasoli$^{1}$, Teague Henry$^{2,3}$,  Heman Shakeri$^{2}$ 
		\footnote{$^1$Department of Mechanical Engineering,
		Payame Noor University, Tehran, Iran. $^2$School of Data Science, University of Virginia, Charlottesville, Virginia. $^3$Department of Psychology, University of Virginia, Charlottesville, Virginia.}}

		\maketitle
	
	\begin{abstract} 
Networks are landmarks of many complex phenomena where interweaving interactions between different agents transform simple local rule-sets into nonlinear emergent behaviors. While some recent studies unveil associations between the network structure and the underlying dynamical process, identifying stochastic nonlinear dynamical processes continues to be an outstanding problem. Here we develop a simple data-driven framework based on operator-theoretic techniques to identify and control stochastic nonlinear dynamics taking place over large-scale networks. The proposed approach requires no prior knowledge of the network structure and identifies the underlying dynamics solely using a collection of two-step snapshots of the states. This data-driven system identification is achieved by using the Koopman operator to find a low dimensional representation of the dynamical patterns that evolve linearly. Further, we use the global linear Koopman model to solve critical control problems by applying to model predictive control (MPC)--typically, a challenging proposition when applied to large networks. We show that our proposed approach tackles this by converting the original nonlinear programming into a more tractable optimization problem that is both convex and with far fewer variables.

	\end{abstract}
	
	\noindent%
  
	\maketitle
	
\section{Introduction}\label{sec:Intro}
Identifying dynamical systems from observations is central to many scientific disciplines, including physical, biological, computer, social sciences, and economics; and opens doors for engineering and interventions \citep{Wang2016Review}. Although inferring the network structure may be impossible in practice, as different networks may display similar dynamical behavior, proper identification of the dynamics remains feasible  \citep{Mieghem2020Predict}. Besides enabling accurate prediction of the network process, the identified model must be well suited for practically implementable strategies to control the underlying dynamics that are both stochastic and nonlinear.  

Deducing laws that govern the relationship between a system's structure and functions is a formidable challenge due to difficulties associated with nonlinearities, stochasticity, high-dimensions, and inherent correlations between the network topology and the underlying dynamical process. As such, approaches such as mean-field approximation \citep{van2009virus, sahneh2013generalized} have been proposed during recent years to model several dynamical processes over networks and offer good estimations under limited circumstances. On the other hand, rapidly developing information technologies leave us with a wealth of data over larger and more diverse networks; further spurring data-driven approaches to construct models without requiring explicit prior knowledge when studying different dynamical processes over networks. 

Our main goal is to establish a data-driven framework for the identification and control of network processes. Overall, a systematic and accurate method for identifying, estimating, and controlling spatiotemporal dynamical features of network processes from data is still an open and challenging issue. To fill this gap, we leverage modern machine learning techniques to model network dynamics in the operator-theoretic setting effectively. This way, we are provided with a purely data-driven approach that assumes no knowledge or identification of network parameters and structure. Furthermore, eigenfunctions of the Koopman operator   summarize the network processes  on low-order manifolds that are evolving linearly. We further establish a tractable framework based on the obtained representation to resolve challenging optimization and control tasks over networks.  Finally, we demonstrate the use of the Koopman operator for system identification and control by applying it to a common model of disease spread on networks, the susceptible-infected-susceptible (SIS) model \citep{van2009virus}. We show that approximating the complex non-linear dynamics of disease spreading using the Koopman operator allows one to develop optimal control regimes that can quickly mitigate the outbreak, a significant improvement over a uniform intervention strategy. Additionally, we show that effective control over infection rates can be accomplished using an appropriately chosen lower-dimensional representation of the high dimensional Koopman operator. 

\subsection{Operator-theoretic approach for the analysis of high-dimensional interconnected system}

Recent advances in data analysis have shown that many complex systems possess dominant low-dimensional invariant subspaces that are hidden in the
high-dimensional ambient space, an underlying structure that enables compact representations for modeling and control \citep{Brunton_Kutz_2019}. To infer these compact representations, operator-theoretic frameworks have been used to address nonlinear relations between subspaces and provide a principled linear embedding for dynamical systems. In particular, the Koopman operator \citep{Mezic2005Spectral} is an infinite-dimensional linear operator that studies the time evolution of measurement functions (observables) of the system state, and its spectral decomposition completely characterizes the behavior of the nonlinear system \citep{Brunton2021ModernKT}. One underlying feature that builds up the Koopman success in the study of nonlinear dynamical systems is its finite-dimensional representation connected with finding effective coordinate transformations in which the nonlinear dynamics appear linear. Stated another way, the Koopman operator explores invariant sets of nonlinear observables that evolve linearly. This is different from conventional approaches that commonly rely on linearization and are only locally valid. As such, the Koopman operator can be viewed as an extension of Hartman–Grobman theorem--which is locally valid within a vicinity of hyperbolic stationary points--into the whole basin of attraction. This offers the prospect of prediction, estimation, and control of nonlinear systems by standard methods developed for linear systems.

The Koopman operator sketches a rich global picture of the nonlinear system  by characterizing several underlying features \citep{Mezic2020Book}. For example, Koopman eigenfunctions at eigenvalue $\lambda=1$ determine the invariant sets, and the eigenfunctions associated with $|\lambda|=1$ form invariant partitions of dynamics \citep{Mezic2005Spectral}. In fact, such eigenfunctions are connected with ergodic and harmonic quotients that reveal coherent structures in dynamics \citep{BudisicQuotient2012}.  Level sets of Koopman eigenfunctions also characterize the sets of points (known as isochrons and isostables) that partition the basin of attraction of limit cycles and fixed points, and reduce such dynamics to action–angle
coordinates \citep{Mauroy2012Isochrons, Mauroy2013Isostables}. \citet{Mezic2016Stability} established relationship between the existence of
specific eigenfunctions of the Koopman operator and the global
stability property of fixed points and limit cycles.
Hence the Koopman operator offers a framework better suited for control by circumventing complexities due to nonlinearity and transforming the nonlinear dynamics into globally linear representations \citep{Proctor2018Control, Brunton2021ModernKT}; e.g. \citet{Brunton2017Chaos} decomposed chaotic systems into intermittently forced linear systems.

In recent years, three main approaches for numerical computation of the Koopman operator are generalized: Laplace analysis, finite section methods, and Krylov subspace methods \citep{Mezic2020Numerical}. Particularly finite section methods construct an approximate operator acting on a finite-dimensional function subspace. The best known such method is dynamic mode decomposition (DMD) \citep{Schmid2010DMD} that features state observables. DMD works based on proper orthogonal decomposition (POD) of high-dimensional linear measurements to extract dynamical patterns that evolve on low-dimensional manifolds \citep{Schmid2010DMD, Tu2014DMD}. Therefore, DMD provides a model in terms of the reduced sets of modes and their progression in time. Although DMD has evolved in recent years into a popular approach to extract linear models from linear measurements \citep{Kutz2016DMDbook}, it inherits the limitations of singular value decomposition and lacks sufficient delicacy to dissect rich nonlinear phenomena and the associated transient dynamics \citep[Chapter 1]{kutz2016dynamic}.

More recently, the extended DMD (EDMD) \citep{Williams2015EDMD} was developed to account for the limitations of DMD, and employs nonlinear observables to recognize a finite-dimensional invariant subset of the Koopman operator that converges to the Galerkin approximation. 
Other variants of DMD are developed to represent different dynamical systems or handle numerical challenges, to mention a few: Kernel-DMD \citep{Williams2015Kernel}, Hankel-DMD \citep{Mezic2017Ergodic}, HAVOK-DMD \citep{Brunton2017Chaos}, tensor-based DMD \citep{Fjii2019GraphDMD}, and recent works that leverage dictionary learning \citep{Li2017Learning} and deep learning architectures \citep{Lusch2018DL, Otto2019Learning, Noe2020DL, Pan2020PhysInfr}.

The simplified representation of complex nonlinear dynamics using the Koopman operator provides exciting opportunities to tackle the challenges in controlling nonlinear systems \citep{Brunton2016Koopman}.
\citet{Mezic2020OptiConstruc} put forward a convex optimization framework for optimal construction of Koopman eigenfunctions
for prediction and control. Several extensions are developed for actuated and controlled systems in \citet{Williams2016Actuated, Kaiser2017Control, Proctor2016Control, Proctor2018Control}. These approaches have recently applied to a wealth of real-world problems  like fluid dynamics \citep{Arbabi2017Fluid, Rowley2017Fluid}, power grids \citep{Mezic2018PowGrid}, molecular dynamics \citep{Noe2018Molecular}, time series classification \citep{Surana2020tSeries}, robotic systems \citep{Abraham2019Active, Bruder2021SoftRobot}, energy consumption in buildings \citep{Mezic2020Bulding, Susuki2020Room}, traffic \citep{Mezic2020Traffic}, spacecraft \citep{Chen2020Attitude}, and hydraulic fracturing operation \citep{Narasingam2020Hydraulic}. 

Given that the Koopman operator's lower-order representation of a complex non-linear system is \textit{linear}, it is very appealing from the perspective of developing control schemes \citep{Mezic2020Book}. Methods for the optimization and on-line control of linear systems are well developed, and the potential to apply these methods productively to the control of complex non-linear systems is a marked advantage of the Koopman operator approach. We want to stress that the Koopman operator captures the dynamics in the whole attraction basin, and thus it can be a more accurate replacement for locally linearized models in these approaches. This is achieved by proper nonlinear measurements in the space of intrinsic coordinates that yield complete information about dynamics. Consequently, the suggested linear predictor is immediately amenable to the range of mature control design techniques, such as optimal control \citep{Brunton2016Koopman,Kaiser2017Eigenf,Huang2018OptContr} or switching control \citep{Sootla2018Pulse}. In particular, since the Koopman works well for short prediction horizons, it is promising for model predictive control (MPC) that needs prediction over a few steps \citep{Mezic2018MPC}. Furthermore, Koopman yields a linear predictor that translates the original nonlinear MPC into a convex optimization problem \citep{Mezic2018MPC} that is more appealing for numerical treatments. 
The Koopman operator has also proven successful for resolving control challenges of partial differential equations (PDEs) by mapping the original nonlinear infinite-dimensional control problem into a low-dimensional linear one \citep{Peitz2019Switch}. Further advantages of this approach in resolving MPC problems can be found in optimizing power grids \citep{Mezic2018PowGrid}, active learning of dynamics for robotic systems control \citep{Abraham2019Active}, and spacecraft altitude stabilization \citep{Chen2020Attitude}.    

\subsection{Spreading Processes on Networks}
Control of dynamical processes over networks is examined recently for mitigation of networked spreading processes. These spreading processes are often used to model the spread of disease through networks that represent person to person contact or interaction, which suggests that effective methods capable of controlling spreading processes have significant public health implications. \citet{Preciado2009Spectral} analyze network spectral properties and consider removing nodes and removing links as control inputs to tame an initial viral infection. In this regard, \citet{Mieghem2011Rmoval} study two problems of optimal node removal and optimal link removal and show them to be NP-complete and NP-hard, respectively and propose greedy strategies based on spectral measures.

Worst-case analysis shows that completely removing nodes or links is ineffective--not to mention node/link removal in real world networks is often impractical, illegal, or both--and latter works considered controlling disease spreading processes by preventive resources and promoting corrective policies. These resources and policies do not alter the structure of the network itself, but rather theoretically modulate the susceptibility to infection of individual nodes and/or the probability that infection will spread along specific links.  \citet{preciado2014optimal} consider both rate-constrained allocation and budget-constrained allocation simultaneously in the framework of geometric programming \citep{preciado2014optimal}. \citet{Nowzari2017Optimal, Watkins2018Optimal, shakeri2015optimal} investigate other variants and provide general solutions. 

The current main strategies for controlling disease spread on networks are to first, allocate resources over the network components (individuals or their ties) to find the minimum required budget to eradicate the disease at the desired rate and second, mitigating the spread in the fastest possible decay rate by allocating a given fixed budget.
The optimization problems have discrete variables, and relaxing them by letting spreading rates take values in a feasible continuous interval aid in numerical solutions. 

One major limitation of the previously described network-based allocation approaches is that they are off-line and thereby without feedback from the current state of the network. This means that these allocation strategies are incapable of adapting to changing demands, leading to at best a non-optimal resource allocation, and at worst a failure to control the disease process due to changing network conditions.  
Optimal control strategies are employed to solve this issue by allowing the control allocation to vary over time \citep{Khanafer2014Optimal, Eshgi2016Malware}; this approach is used in \citet{Kandhway2016Information, Mieghem2019Optimal} for application in virus spreading problems and \citet{Dashtbali2020Game} for investigating social distancing in response to epidemic using differential games approach. 

\citet{Watkins2020MPC} use MPC for optimal containment of epidemic over networks. In particular, and during the recent outbreak of COVID-19, the significance of identifying and intercepting the virus spread over networks is more evident.  \citet{Carli2020Covid19} study mitigating the outbreak using a multi-region scenario, with the underlying network representing inter-region mobility and propose a model-based MPC where the parameters of the model are fitted based on the collected data from the network of Italian regions. 

Despite the vast literature, finding practical approaches for controlling epidemics over complex networks remains an outstanding problem with real-world assumptions and the corresponding uncertainties and unknowns that pose challenges for model-based approaches.
We present a summary of the main results here. Firstly, the existing control methods are specialized for deterministic models often approximated from the original stochastic models. Despite establishing connections between the two, these connections are only relevant for simplified cases, and additionally, the connections between control solutions of the two models are unclear. Second, the current approaches admit centralized solutions of computational burden, making them intractable for large networks. Third, conventional methods assume no uncertainties and require complete knowledge of everything, including natural recovery rates, infection rates, state information, and network topologies. 
Avoiding the above simplifying assumptions while having tools that can handle network and parameter variations are necessary for practical approaches. 

\subsection{Contribution statement}
This work attempts to address the shortcomings of the available approaches discussed above, where we consider identifying and controlling epidemics solely based on our spatio-temporal observations. Our approach is intended to tackle identifying and control of stochastic processes over complex networks using several features; First, our method is purely data-driven and have no assumptions about network parameters, structure, or the underlying dynamical process, and  is based on the original stochastic process that produced the data at the outset. Second, to reach an effective data-driven method that is tractable for optimization and control over large networks, we use the latest achievements in machine learning and operator-theoretic to identify a Koopman representation that is \textit{interpretable}, \textit{low-dimensional}, and \textit{linear}. This way, the underlying high-dimensional dynamics is represented through extracting the most effective modes evolving linearly under the networked processes. Third, we revisit the important MPC problem over complex networks and show our proposed approach maps the original high-dimensional nonlinear optimization problem into a low-dimensional convex representation that is well suited for existing numerical approaches and enables real-time softwares. 

We organize the paper as follows. Section \ref{sec:Koopman} presents the data-driven Koopman identification of stochastic processes over networks. Section \ref{sec:MPC} explains the nonlinear MPC problem over networks and its transformation into Koopman MPC. In Section \ref{sec:results}, we apply our approach to a Markov process representing the SIS epidemic model over network. Section \ref{sec:conclusion} is devoted to concluding remarks and discussion.

\section{Koopman operator}\label{sec:Koopman}
\subsection{Problem statement}     
We consider a controlled discrete-time Markov process taking place over a network as
\begin{equation}\label{eq:model}
\begin{gathered}
x\mapsto F(x,u;\omega)
\end{gathered}
\end{equation}
where $x\in\mathcal M\subseteq\mathbb R^n$ is the system state vector in the state space $\mathcal M$, $u\in\mathcal U\subseteq\mathbb R^l$ is the control input, and $\omega\in\Omega$ is an element in the probability space associated with the stochastic dynamics $\Omega$ and probability measure $P$. The aim is to obtain a nonlinear embedding mapping (transformation)  $\psi=[\psi_1,...,\psi_N]^T$, with $\psi_i:\mathcal M\longrightarrow \mathbb R$, from the original state space $\mathcal M$ to a subset of $\mathbb R^N$ that enables us to construct a linear predictor for the expected value $\mathbb E[x(k)]$  of the form
\begin{equation}\label{eq:predictor}
\begin{gathered}
z(k+1)=Az(k)+Bu(k) \\
\mathbb E[\hat x(k)|\hat x(k-1)]=Cz(k) \\
z_0=\psi(x_0)
\end{gathered}
\end{equation}
where $A\in\mathbb R^{N\times N}$, $B\in\mathbb R^{N\times l}$, $C\in\mathbb R^{n\times N}$, and the output $\mathbb E[\hat x(k)|\hat x(k-1)]$ is used for prediction and control of the expected value of the state vector $x$ given the initial condition $x_0$. 
Therefore, we seek a linear system that faithfully represents the original nonlinear
dynamical system. The key is
finding proper transformation $\psi$ that maps the original state $x\in\mathcal M\subseteq\mathbb R^n$ to the lifted state $z\in\mathbb R^N$, with (typically) $N\gg n$, that evolves linearly--though the number of dimensions $N$ is still of concern for practical implementation, we will postpone the discussion on reducing the dimensionality to Section \ref{sec:RedcedKoopman}. The predictor model \eqref{eq:predictor} is amenable to linear control design approaches in the lifted space $z$. 
Moreover, $u$ remains unlifted in \eqref{eq:predictor} allowing direct use of linear constraints on input or/and states  in the lifted state. As long as the predictions of \eqref{eq:predictor} are accurate for short time horizons, it is desirable for the use of linear control methodologies (such as linear MPC).

\begin{figure*}
	\centering
	\subfloat[\label{fig:Identification}]{\includegraphics[clip,width=.9\columnwidth]{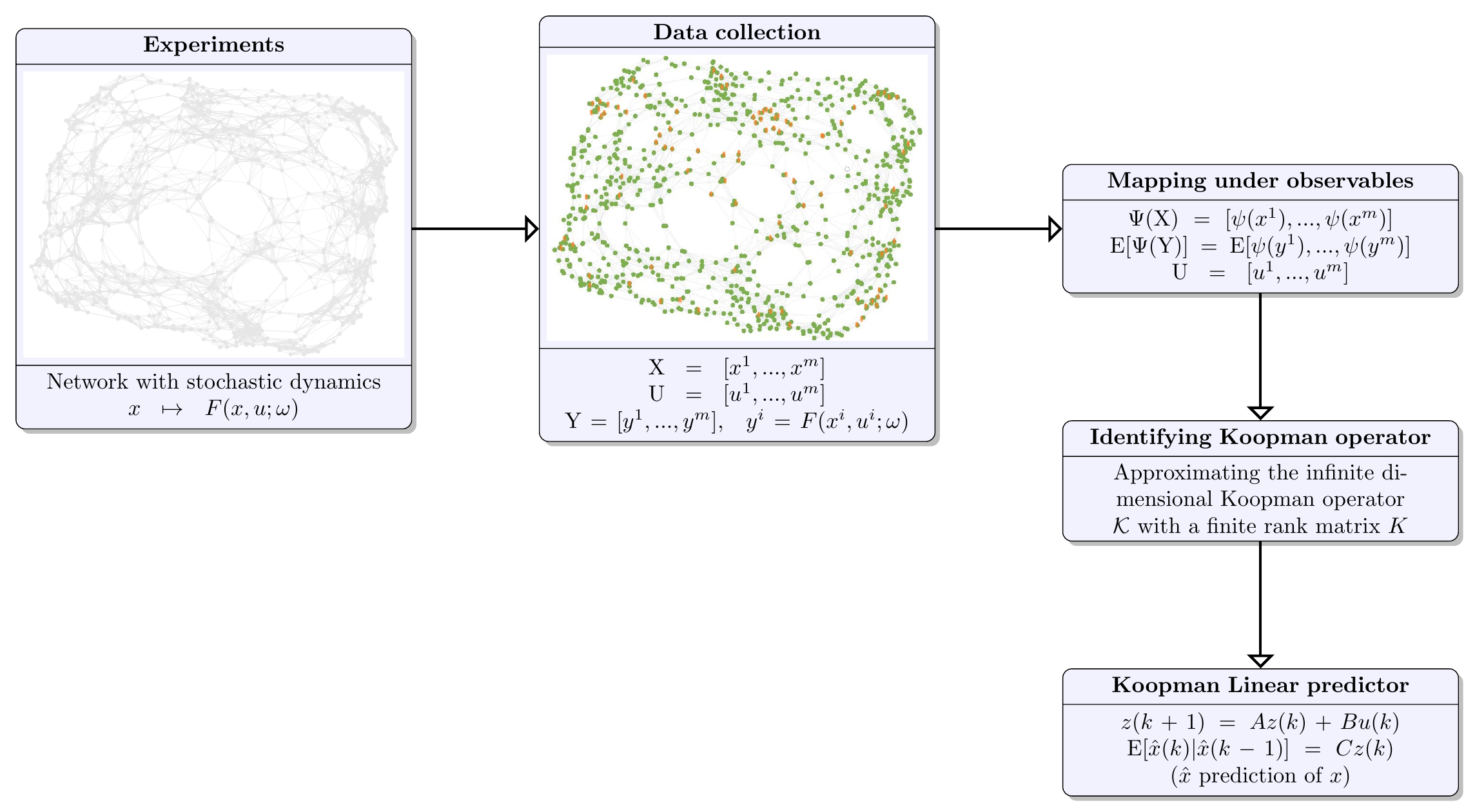}} \\
	\subfloat[\label{fig:MPC}]{\includegraphics[clip,width=.7\columnwidth]{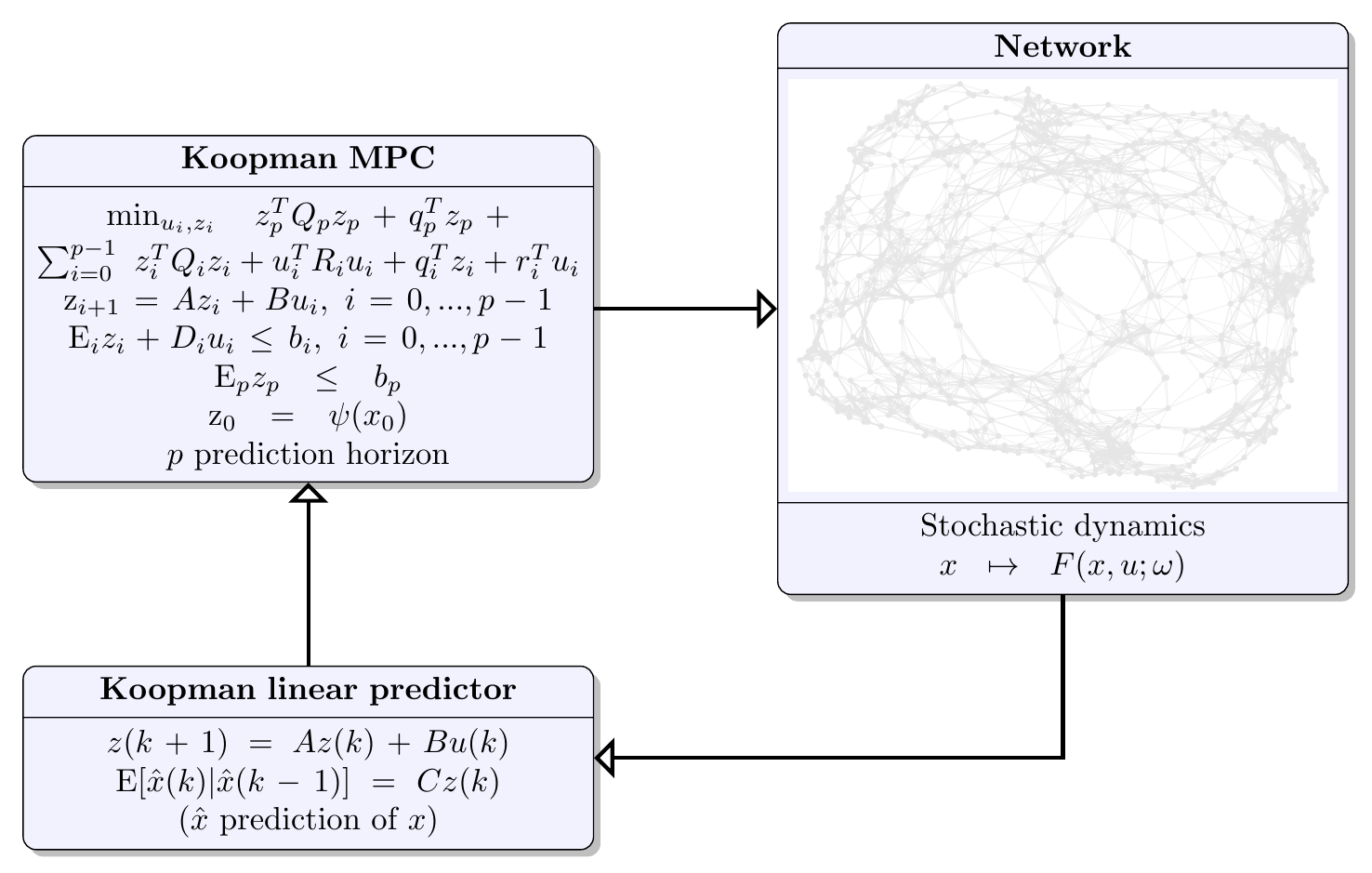}}
	
	\caption{Main idea. (a) Koopman identification. We only collect stochastic (binary) data in the form of $m$ snapshots including $(x^i, u^i, y^i)$ with $y^i=F(x^i,u^i;\omega)$. Although we assume no prior knowledge of the system parameters, the underlying dynamics, or network geometry, one can still incorporate available information to enrich the proposed analysis. The Koopman operator is an infinite-dimensional operator that specifies how functions of state evolve, so that it is projected into an (invariant) observables subspace, where the lifting set $\psi(x)$ mapps the original system into a linear one in higher dimension. For effective low-order modeling, a Koopman mode decomposition represents the dynamics in the most effective Koopman modes. (b) The Koopman operator translates the original nonlinear MPC into a convex problem that is amenable for numerical solution. We further use the reduced-order Koopman to decrease the size of the optimization. }
	\label{fig:main} 
\end{figure*}

Next we demonstrate that such transformation can be established in the framework of Koopman operator by using data of the triplet form $x(k),~x(k+1),~u(k)$ (see Figure \ref{fig:main} for a sketch of the main ideas in the paper).

\subsection{Koopman operator for controlled stochastic processes}
Koopman operator embeds the nonlinear dynamics into an appropriate Hilbert space $\mathcal F$ where the dynamics is linear and one can construct the predictor \eqref{eq:predictor}. Namely, Koopman is a linear operator of infinite dimension that takes a scalar observable-function $g: \mathcal M \times \mathcal U\times \Omega\longrightarrow\mathbb R$ belonging to $\mathcal F$ and gives its expected value evolution in the state space. The function space $\mathcal F$ is invariant under the action of the Koopman operator. Additionally, to fully describe the underlying dynamical system, $\mathcal F$ must contain the components of the state vector $x$. 

Following \citet{Proctor2018Control}, who generalize the Koopman operator for systems with input, we consider the Koopman operator $\mathcal K:\mathcal F\longrightarrow\mathcal F$ for stochastic process \eqref{eq:model} as
\begin{equation}\label{eq:Koopman2}
\begin{gathered}
\left[\mathcal Kg\right](x_k,u_k;\omega)=\mathbb E[g(F(x_k,u_k;\omega),u_{k+1};\omega)]=\mathbb E[g(x_{k+1},u_{k+1};\omega)]
\end{gathered}
\end{equation}
with $x_k=x(k)$ and $u_k=u(k)$. 
Including actuation in \eqref{eq:Koopman2} renders the Koopman family non-autonomous. In the equivalent autonomous formulation, \citet{Mezic2018MPC} extend the system state to include all control sequences and apply the shift operator to advance the observation of input.
The Koopman operator's spectral properties are directly connected with several geometric characteristics, e.g., invariant sets and partitions \citep{Mezic2005Spectral} or asymptotic behavior \citep{Mauroy2012Isochrons, Mauroy2013Isostables}, of the underlying nonlinear system. Moreover, the Koopman modes associated with Koopman eigenfunctions can yield the evolution of observables and the orbits of the system for all initial conditions. In this regard, the Koopman operator gives a complete description of the underlying nonlinear system, provided that the space of observables $\mathcal F$ spans the elements of $x$. If $\varphi\in \mathcal F$ is an eigenfunction of Koopman operator and $\lambda$ its eigenvalue, then the spectral problem of Koopman operator reads $\mathcal K\varphi=\lambda \varphi$. If $\varphi_1, \varphi_2\in\mathcal F$ are eigenfunctions of $\mathcal K$ with eigenvalues $\lambda_1$ and $\lambda_2$, then $\varphi_1\varphi_2$ is eigenfunction of $\mathcal K$ with eigenvalue $\lambda_1\lambda_2$. This is an implication of the Koopman operator being (generally) infinite-dimensional.  \citet{Mezic2012Koopmanism} and \citet[Chapter~1]{Mezic2020Book} provide a detailed review of Koopman operator properties. 

The infinite-dimensional Koopman operator $\mathcal K$ is approximated by its finite-dimensional projection $K$ using data-driven approaches that are well suited for this purpose. DMD provides a projection of Koopman operator onto the space of linear observables   \citep{Brunton_Kutz_2019} and  EDMD  produces more precise approximations by incorporating nonlinear observables that result in a higher-dimensional approximation (see \citet{Williams2015EDMD, Mauroy2019Lifting} for deterministic and \citet{Noe2020Markov} for stochastic systems). Using an extended state vector for systems including input and with the shift operator of a known input profile, \citet{Mezic2018MPC} argue that finite-dimensional approximation $K$ to the
operator $\mathcal K$ yields a predictor of the form \eqref{eq:predictor}. Specifically, $K$ is the projection of $\mathcal K$ onto a subspace $\bar{\mathcal F}\subseteq\mathcal F$ of observables spanned by $( \psi_i(x), u)$, where the lifting functions $\psi_i, i=1,...,N,$ only act on the state $x$ and the control input $u\in\mathcal U$ remains unlifted\footnote{If subspace $\bar{\mathcal F}$ is invariant under $\mathcal K$, then all of the
eigenvalues and eigenfunctions of $K$ are also eigenvalues and eigenfunctions of $\mathcal K$ \citep{Otto&Rowley2021Review}. However, since usually such an invariant subspace $\bar{\mathcal F}$ is not known in advance, with $K$ we obtain an approximation for $\mathcal K$.}. As such, the control input appears linearly in the resulting model, which is amenable for control design purposes.  

\subsection{Finite-dimensional projection using EDMD with control}
Recent uses of Koopman operator  in control architecture are focused in the context of deterministic systems \citep{Proctor2018Control, Mezic2018MPC, Peitz2019Switch, Brunton2021ModernKT}; in particular, we follow \citet{Mezic2018MPC} and assume that the data is collected in the form of $m$ snapshots as
\begin{equation*}
\begin{gathered}
\text X=[x^1, ..., x^m], \ \ \ \text Y=[y^1, ..., y^m], \ \ \ \text U=[u^1, ..., u^m], \ \ x^i, y^i\in\mathbb R^n, \ \ u^i\in\mathbb R^l 
\end{gathered}
\end{equation*} 
where $y^i=F(x^i,u^i;\omega)$. Unlike the original DMD formulation \citep{Rowley2009, Schmid2010DMD}, the data need not be sequentially ordered along a single trajectory of \eqref{eq:model} as $y^i=x^{(i+1)}$, and we generally use different snapshot triples $(x^i,y^i,u^i)$ along different trajectories (corresponding to different initial conditions with generally $y^i\neq x^{i+1}$). The action of lifting functions is then given as 
 \begin{equation*}
 \begin{gathered}
 \Psi(\text X)=[\psi(x^1), ..., \psi(x^m)]
 \end{gathered}
 \end{equation*}
where $\psi(x)=[\psi_1(x), ..., \psi_N(x)]^T$ is a given dictionary of nonlinear functions. For stochastic processes, we estimate the expected value of $\Psi(\text Y)$ directly from experiments.
\begin{equation*}
\begin{gathered}
\mathbb E[\Psi(\text Y)]=[\mathbb E[\psi(y^1)], ..., \mathbb E[\psi(y^m)]]
\end{gathered}
\end{equation*}
Then the matrices $A,B,C$ in \eqref{eq:predictor} are solutions of following optimization problems:
\begin{equation}\label{eq:opt1}
\begin{gathered}
\min_{A,B}{\|\mathbb E[\Psi(\text Y)]-A\Psi(\text X)-B\text U\|} \\
\end{gathered}
\end{equation}
\begin{equation}\label{eq:opt2}
\begin{gathered}
\min_{C}{\|\text X-C\Psi(\text X)\|}
\end{gathered}
\end{equation}
We solve \ref{eq:opt1} and \ref{eq:opt2} using the normal equations (see \citet{Mezic2018MPC} for a discussion on the numerical considerations).

\subsection{Reduced-order linear representation}\label{sec:RedcedKoopman}
Koopman operator embeds the networked nonlinear dynamical system into a linear system but with a higher dimension.  One desires a low-dimensional model in practice for fast optimization and real-time control.
In the context of linear measurements, the basic DMD scheme extends to include exogenous effects and uses a truncated set of decomposed low-energy modes for order reduction \citep{Proctor2016Control}. Here, we develop this approach to Koopman mode decomposition and establish reduced-order Koopman representation with control. For this purpose,  we start by a singular value decomposition    
\begin{equation}
    [\Psi(X)^T \ \  U^T]^T=U_1\Sigma_1V_1^*
\end{equation}
where $U_1$ is  bipartite, i.e., $U_1=[U_{11}^T \ \ U_{12}^T]^T$ based on the model dimensions.  Second, we perform SVD on
\begin{equation}
    \mathbb E[\Psi(Y)]=U_2\Sigma_2V_2^*,
\end{equation}
where the truncation value is $r$; hence, a reduced Koopman model of order $r$ is established. The low-dimensional model matrices are computed as
\begin{equation}\label{eq:ReducMatrix}
\begin{gathered}
\tilde A=U_2^T\mathbb E[\Psi(Y)]V_1\Sigma_1^{-1}U_{11}^TU_2 \\
\tilde B=U_2^T\mathbb E[\Psi(Y)]V_1\Sigma_1^{-1}U_{12}  \\
\tilde C=CU_2
\end{gathered}
\end{equation}
Thus the Koopman model \eqref{eq:predictor} reduces to the coordinate $z=U_2^T\psi(x)$ by replacing $A$, $B$, and $C$ with $\tilde A$, $\tilde B$, and $\tilde C$. In this regard, we use the first $r$ Koopman modes to construct a low-dimensional network process representation summarized in Algorithm \ref{alg:KI}. This strategy's success lies in the existence of a low-dimensional manifold on which the underlying dynamics evolve. Although this manifold depends on the control input, we will illustrate that this approach is sufficiently powerful to effectively capture manifolds for a given input training range. In other words, while our observations are in high-dimension over networks, the actual collective dynamics evolve in low-dimension. The accuracy of this manifold identification improves with narrowing the input training range. \citet{Peitz2019Switch} partition the input space into a set of subspaces and extract a surrogate model for each range; though the combinatorial nature of this approach prohibits its use on high-dimensional input spaces present in networks. 

\begin{figure}
    \begin{minipage}{\linewidth}
        \begin{algorithm}[H]
           \caption{Reduced Koopman identification of networked dynamics with inputs}
           \label{alg:KI}
           Inputs: Data matrices $X$, $U$, $\Psi(X)$, and $\mathbb E[\Psi(Y)]$ \\
           Outputs: Koopman model matrices $\tilde A$, $\tilde B$, $\tilde C$
           \begin{algorithmic}[1]
                   \State Choose a truncation value $r$
                   \State SVD: $[\Psi(X)^T \ \  U^T]^T=U_1\Sigma_1V_1^*$
                   \State Use the number of observables $N$ to bipartite $U_1=[U_{11}^T \ \ U_{12}^T]^T$ 
                   \State SVD: $\mathbb E[\Psi(Y)]=U_2\Sigma_2V_2^*$ and truncate it for first $r$ modes
                   \State Solve \eqref{eq:opt2} to get $C$
                   \State $\tilde A \gets U_2^T\mathbb E[\Psi(Y)]V_1\Sigma_1^{-1}U_{11}^TU_2$, $\tilde B \gets U_2^T\mathbb E[\Psi(Y)]V_1\Sigma_1^{-1}U_{12}$, $\tilde C \gets CU_2$
           \end{algorithmic}
        \end{algorithm}
    \end{minipage}    
\end{figure}

\subsection{Choosing an appropriate subset in the function space}\label{sec:functions}
Under the assumption of sufficiently rich basis and a large number of functions, one can expect a small approximation error \citep{Williams2015EDMD}. However, it is an open question: what type of observables will yield the best result for a specific problem. There are three popular choices: Hermite polynomials, radial basis functions (RBFs), and discontinuous spectral elements \citep{Williams2015EDMD}. A partially optimized space of observables can be attained by first selecting a parameterized feature space \citep{Noe2020Markov}, e.g. Gaussian RBFs parameterized with the smoothing parameter, and then optimize the associated parameters (the smoothing parameter in case of Gaussian RBFs). Recent investigations of dictionary learning representation by \citet{Li2017Learning, Yeung2019Learn, Otto2019Learning} are extremely promising. Generally, the physics of the problem, e.g., continuity property and locality, can also be used in determining the choice and the number of basis functions \citep{Chen2019Complexity}. 

The dictionary could also include the system state observable (see, e.g., \citet{Williams2015EDMD, Mezic2018MPC}). This will enhance the linear state reconstruction from observables, i.e., decoding back to the original coordinates. However, it requires at least as many functions as the dimension of the original system state, which  is undesirable for large networks  evolving in lower intrinsic dimensions. Furthermore, the linear state observable generally lacks high enough resolution to capture complex features of nonlinear systems. Hence, when the full state observable is absent in the Koopman eigenfunction set, forcing the full state observable constraint in the Koopman-invariant subspace will result in overfitting. Moreover, it is impossible to determine a finite-dimensional Koopman-invariant subspace that includes the original state variables for any system with multiple
fixed points or any more general attractors \citep{Brunton2016Koopman}.

\section{Model predictive control for networked processes}\label{sec:MPC}
The fundamental idea behind the MPC is to measure the current state and design an open-loop optimal control over a finite-time horizon based on a predictive model. For a closed-loop control behavior, the MPC applies only the first portion of the synthesized control during a short time interval. The controller uses the updated state measurements to design the next open-loop control function--and this procedure repeats in the subsequent steps. Therefore MPC yields a closed-loop control approach that concurrently optimizes system performance, handles nonlinearity, holds robustness properties, incorporate input and state constraints with desirable (stability) convergence properties--we refer the reader to \citet{Grune2017MPC} for an exposition. Extension to stochastic systems and the cumulative reasons above lead to the fast growth of the MPC paradigm in the control systems literature. 

\subsection{Original MPC}
For the original stochastic process in \eqref{eq:model}, we consider a  nonlinear MPC problem that at each time step of the closed-loop operation solves the following optimization problem
\begin{equation}\label{eq:MPC}
\begin{gathered}
\min_{u_i,\mathbb E[\bar x_i]} \ \ \ l_p(\mathbb E[\bar x_p])+\sum_{i=0}^{p-1} \ l_i(\mathbb E[\bar x_i])+u_i^TR_iu_i+r_i^Tu_i \\
\text{subject to} \ \ \ \bar x_{i+1}=F(\bar x_i,u_i;\omega), \ \ \ \ \ i=0, ...,p-1 \\
\ \ \ \ \ \ \ \ \ \ \ \ \ \ \ \ \ \  C_i(\mathbb E[\bar x_i])+D_iu_i\leq b_i, \ \ \ \ \ i=0, ...,p-1 \\
C_p(\mathbb E[\bar x_p])\leq b_p \\
\bar x_0=x_0
\end{gathered}
\end{equation}
where $x_0$ is the current state, $p$ is the prediction horizon, $\mathbb E[\bar x]$ denotes the prediction of $\mathbb E[x]$, $l_i$ is nonlinear scalar valued and $C_i\in\mathbb R^{n_c}$ nonlinear vector valued functions of state vector expected value, $R_i\in\mathbb R^{l\times l}$ is positive semidefinite, vector $r_i\in\mathbb R^{l}$, vector $b_i\in \mathbb R^{n_c}$, and matrix $D_i\in\mathbb R^{n_c\times l}$, with $n_c$ the number of constraints. At each time step, only the first element of the optimal control sequence is applied and the optimization is repeated in the next time step. 

The optimization problem \eqref{eq:MPC} is, in general, nonconvex and hard to solve to achieve global optimality, particularly for large networks. Furthermore, we have generally no prior realization of the dynamics $F(.)$ for accurate state prediction. Applying the Koopman operator, we transform this problem into a low-order convex optimization problem that is numerically tractable.

\subsection{MPC via Koopman}
The Koopman operator transforms the original MPC problem \eqref{eq:MPC} into the following convex problem
\begin{equation}\label{eq:KoopmanMPC}
\begin{gathered}
\min_{u_i, z_i} \ \ \ z_p^TQ_pz_p+q_p^Tz_p+\sum_{i=0}^{p-1} \ z_i^TQ_iz_i+u_i^TR_iu_i+q_i^Tz_i+r_i^Tu_i \\
\text{subject to} \ \ \ z_{i+1}=Az_i+Bu_i, \ \ \ \ \ i=0, ...,p-1 \\
\ \ \ \ \ \ \ \ \ \ \ \ \ \ \ \ \ \  E_iz_i+D_iu_i\leq b_i, \ \ \ \ \ i=0, ...,p-1 \\
E_pz_p\leq b_p \\
z_0=\psi(x_0)
\end{gathered}
\end{equation}
where $Q_i\in\mathbb R^{N\times N}$ is positive semidefinite and $q_i\in\mathbb R^N$. The matrices $E_i\in\mathbb R^{n_c\times N}$ define the state constraints, which become linear in lifted space. The optimization problem \eqref{eq:KoopmanMPC} is  convex, i.e, quadratic programming. 

Suggested by \citet{Mezic2018MPC}, one can transform the original optimization problem \eqref{eq:MPC} into  \eqref{eq:KoopmanMPC} by constructing the matrices $A$ and $B$ and the vector $z_0$ using the $\psi(.)$ embeddings (see Section \ref{sec:Koopman}) with including in the lifting set the functions $\psi_{i+1}(x)=l_i(x)$ and $\psi_{(p+in_c+2:p+(i+1)n_c+1)}=C_i(x)$ for $i=0, \cdots,p$. Consequently
$$Q_i=0, \text{~and ~}q_i=[\boldsymbol 0_{1\times i}, 1, \boldsymbol 0_{ 1\times (N-i-1)}],$$
$$E_i=[\boldsymbol 0_{n_c\times (p+in_c+1)},I_{n_c\times n_c},\boldsymbol 0_{n_c\times(N-p-(i+1)n_c-1)}],$$
where $\boldsymbol 0_{i\times j}$, $\boldsymbol 1_{i\times j}$, and $I_{i\times j}$ are all zeros, all ones, and identity matrices, respectively. 
Although this canonical approach always returns a linear cost function, if $l_i(x_i)$ is quadratic, we opt for the freedom of  \eqref{eq:KoopmanMPC} and instead of setting $\psi_i=l_i$, use the Koopman output matrix $C$ to consider quadratic terms in the cost function of \eqref{eq:KoopmanMPC}, thereby reducing the dimension of the lift. 

\citet{Mezic2018MPC} show that the computational complexity of solving the MPC problem \eqref{eq:KoopmanMPC} can be rendered independent of the dimension of the lifted state $N$ by transforming to a dense form. Hence, the computational cost of solving the dense form is comparable to solving a standard linear MPC on the same prediction horizon, with the same number of control inputs and the state space's dimension equal to $n$ rather than $N$. Although we follow this strategy in our numerical programmings for the full-order Koopman MPC, we are less concerned about the dimensionality when using the reduced-order Koopman for MPC since our approach uses a low-dimensional model.  Recall that the first $r$ modes $U_2^T\psi(x)$ are used to represent the dynamics in the reduced-order Koopman MPC framework. Therefore,  matrices $A, B, C$ in \eqref{eq:KoopmanMPC} are replaced with $\tilde A,\tilde B,\tilde C$, respectively, and the dimension value $N$ is replaced with $r\ll n<N$. Algorithm \ref{alg:MPC} shows the reduced-order Koopman MPC procedure when each function $l_i(\mathbb E[x])$ in the original MPC problem \eqref{eq:MPC} is quadratic in terms of state vector expected value through the positive definite matrix $\hat Q_i\in\mathbb R^{n\times n}$ and vector $\hat q_i\in\mathbb R^n$, and each $C_i(\mathbb E[x])$ is linear through matrix $\hat E_i\in\mathbb R^{n_c\times n}$.

\begin{figure}
    \begin{minipage}{\linewidth}
        \begin{algorithm}[H]
           \caption{Network MPC via Koopman}
           \label{alg:MPC}
           Cost function in \eqref{eq:MPC}: $\mathbb E[\bar x_p]^T\hat Q_p\mathbb E[\bar x_p]+\hat q_p^T\mathbb E[\bar x_p]+\sum_{i=0}^{p-1} \ \mathbb E[\bar x_i]^T\hat Q_i\mathbb E[\bar x_i]+u_i^TR_iu_i+\hat q_i^T\mathbb E[\bar x_i]+r_i^Tu_i$ \\
           Constraints in \eqref{eq:MPC}: $\hat E_i\mathbb E[\bar x_i]+D_iu_i\leq b_i \ \text{for} \ i=0, ...,p-1, \hat E_p\mathbb E[\bar x_p]\leq b_p$ \\
           Input: Current system state $x_0$\\
           Output: Control input 
           \begin{algorithmic}[1]
                   \State For $i=0,...,p$, set $q_i=\tilde C^T\hat q_i$, $Q_i=\tilde C^T\hat Q_i\tilde C$, $E_i=\hat E_i\tilde C$
                   \State Solve the convex optimization \eqref{eq:KoopmanMPC} for $A=\tilde A$ and $B=\tilde B$
                   \State Only keep and apply the first computed control input $u_0$ 
                   \State Update the current system state $x_0$ and repeat the procedure for the next time step
           \end{algorithmic}
        \end{algorithm}
    \end{minipage}    
\end{figure}

\section{Application: network SIS epidemic model}\label{sec:results}
We apply our proposed approach to study the networked SIS model--a benchmark to study epidemics over networks. We give a short description of the SIS model in the next subsection but encourage the reader to see \citet{van2009virus} for a detailed study of dynamical properties.

\subsection{Underlying Markov process}
\begin{figure}[h]
	\centering
	\includegraphics[width=0.3\textwidth]{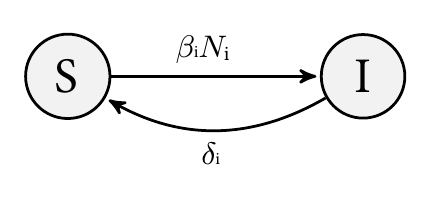}
	\caption{Transition graph for node $i$ with $N_i$ number of infected neighbors in the SIS model.}
	\label{fig:SIS}
\end{figure}

The Markov process is defined based on a set of rules describing the possible transitions between different compartments. In the standard network SIS model \citep{van2009virus}, a susceptible agent $i$ adjacent to an infected neighbor experiences infection through a Poisson process with the rate $\beta_i$--the independent processes merge, and thus the infection rate increases with the number of infected neighbors. 
 Similarly, an infected agent $i$  recovers back to the susceptible state with a Poisson process with the rate $\delta_i$. Figure \ref{fig:SIS} shows the transition diagram where $S$ and $I$ denote the Susceptible and Infected compartments respectively, and $N_i$ denotes the number of infected agents neighboring agent $i$. 
 
 For each node $i\in \{1,...,n\}$, consider a binary random variable $X_i$, and denote $X_i^t$ the value of $X_i$ at time $t$, i.e., $X_i^t\in\{S, I\}$. The  transitions between  \textit{S} and \textit{I} are modeled via the following continuous-time Markov process:

\begin{equation}\label{eq:SIS}
\begin{gathered}
\text{Pr}\left(X_i^{t+\Delta t}=I|X_i^t=S, \boldsymbol{X}^t\right)=\beta_i  N_i^t \Delta t \\
\text{Pr}\left(X_i^{t+\Delta t}=S|X_i^t=I, \boldsymbol{X}^t\right)=\delta_i \Delta t+o\left(\Delta t\right) 
\end{gathered}
\end{equation}
where   $\boldsymbol{X}^t\stackrel{\Delta}{=}\{X_i^t, i=1,...,n\}$ is the joint state of the network, $N_i^t$ is the value of $N_i$ at time $t$, and $\Delta t$ is the time step that undergoes a Poisson process. 

\subsection{Koopman identification}
We use the stochastic approach proposed in GEMF by \citet{Darabi2017GEMF} to simulate the SIS Markov process \eqref{eq:SIS} on arbitrary networks. At each time step, the state vector $x\in\mathbb R^n$ is a discrete binary vector, where the $i$-th element of $x$ is $0$ if agent $i$ is susceptible and $1$ if infected. 
Algorithm \ref{alg:data} describes the  data generation and aggregation using  stochastic simulators. We choose a number of $n_{traj}$ initial conditions randomly initiated from $[0,1]^n$. For each fixed initial condition, we then simulate \eqref{eq:SIS} for $n_{sim}$ times and average to obtain the expected values of dictionary functions $\psi$. 
We record the first and last data of each simulation running for the time period $T$. Therefore, the mapping \eqref{eq:model} takes $x(t)$ and gives $x(t+T)$. To learn the system response to a range of inputs, we select a random perturbation vector $U$ within a given range $[\underline u,\bar u]$ and apply that input throughout the corresponding trajectory. 

\begin{figure}
	\begin{minipage}{\linewidth}
		\begin{algorithm}[H]
			\caption{Data generation and collection }
			\label{alg:data}
			Inputs: $n_{traj}=m$, $n_{sim}$, $T$, $\underline u$, $\bar u$ \\
			Outputs: Data matrices $X$, $U$, and $\mathbb E[\Psi(Y)]$ \\
			GEMF simulator takes the current state $x(t)$ and the picewise-constant control input $u(t)$ and gives the network state $x(t+T)$ at $t+T$
			\begin{algorithmic}[1]
			    \For{$i=1:n_{traj}$} 
			        \State Randomly generate $x^i$ in $[0, 1]^n$ including 0 and 1 elements
			        \State Randomly generate $u^i$ in $\mathbb R^n$ satisfying $\underline u\leq u\leq \bar u$
			        \State $\mathbb E[\psi(y^i)] \gets 0$
			        \For{$j=1:n_{sim}$}
			            \State Run the GEMF for $x(0)=x^i$ and $u^i$ and get $y^{ij}=x(T)$
			            \State Compute $\psi(y^{ij})$
			            \State $\mathbb E[\psi(y^i)] \gets \mathbb E[\psi(y^i)]+\psi(y^{ij})/n_{sim}$
			        \EndFor
			     \EndFor 
                 \State $X \gets [x^1,...,x^{n_{traj}}]$, $U \gets [u^1,...,u^{n_{traj}}]$, $\mathbb E[\Psi(Y)] \gets [\mathbb E[\psi(y^1)],...,E[\psi(y^{n_{traj}})]]$
			\end{algorithmic}
		\end{algorithm}
	\end{minipage}
\end{figure}

We consider the constant function 1 and Gaussian radial basis functions (RBF) for the dictionary functions. We choose the RBF centers from $k$-means clustering \citep{Bishop2006} with a pre-specified value of $k$ on the combined data set. Doing so, the RBF centers are directly related to the density of data points, effectively distributing the RBF centers over the cloud of points  \citep{Williams2015EDMD}. 

We adopt the variation of infection rates $\beta_i$, $i=1,...,n$, as inputs to the spreading dynamics, letting $\beta_i={\beta_0}_i-\Delta\beta_i$ with ${\beta_0}_i$ indicating a constant (passive) infection rate and $u_i=\Delta\beta_i$ the input to agent $i$. In practice, the infection rate can be regulated by putting restrictions on traffic/travel,  quarantining subpopulation, distributing masks, vaccinations, or raising awareness about the disease \citep{Nowzari2016Review}. 
The control input $u_i$ is constrained by constants $\bar u$ and ${\beta_0}_i$ as $0\leq \bar u\leq u\leq {\beta_0}_i$; thus the total infection rate $\beta_i$ remains nonnegative. One may also constraint the total control input for all agents by $u_T$ as $\sum_{i=1}^{n}u_i\leq u_T\leq \sum_{i=1}^{n}{\beta_0}_i$.

We considered and examined our approach on three random graph models: randomly generated geometric (Geo), Erd\H{o}s-R\'enyi (ER), and Watts-Strogatz (WS) graphs as testbeds each with $n=100$ nodes and a fixed average degree $\hat{d} = 10$. 
To conserve space, whenever the results of other models can be interpreted similarly, we present only the results for ER networks. We compare our data-driven approach in predicting the networked dynamics to the epidemic mean-field model at which we provide both graph structure and the SIS model \citep{sahneh2013generalized}. Note that we are unable to offer a similar comparison for the control of networked processes (Section \ref{sec:MPC}) due to lack of known algorithms able to handle large-scale graphs--even with the knowledge of network structure and nodal dynamics. 

We set ${\beta_0}_i=1$, $\delta_i=2$, $n_{traj}=2\times10^4$, $n_{sim}=10$, and $T=1$  in Algorithm \ref{alg:data}.  Moreover, we consider averaging the prediction error over $1000$ randomly chosen initial conditions that are allowed to evolve for a time period $t=T$. Although this time period is equivalent to one  step in Equation \eqref{eq:predictor}, i.e. the operator sense, it includes multiple transitions (events) in Equation \eqref{eq:SIS}, i.e. GEMF stochastic simulator. Furthermore, choosing a large $T$ may incorporate less of the transient pass and even result in better metrics of prediction, but it lacks precision for our control design later.

\subsubsection{Constant input}

\begin{figure}
	\centering
	\subfloat{\includegraphics[clip,width=.33\columnwidth]{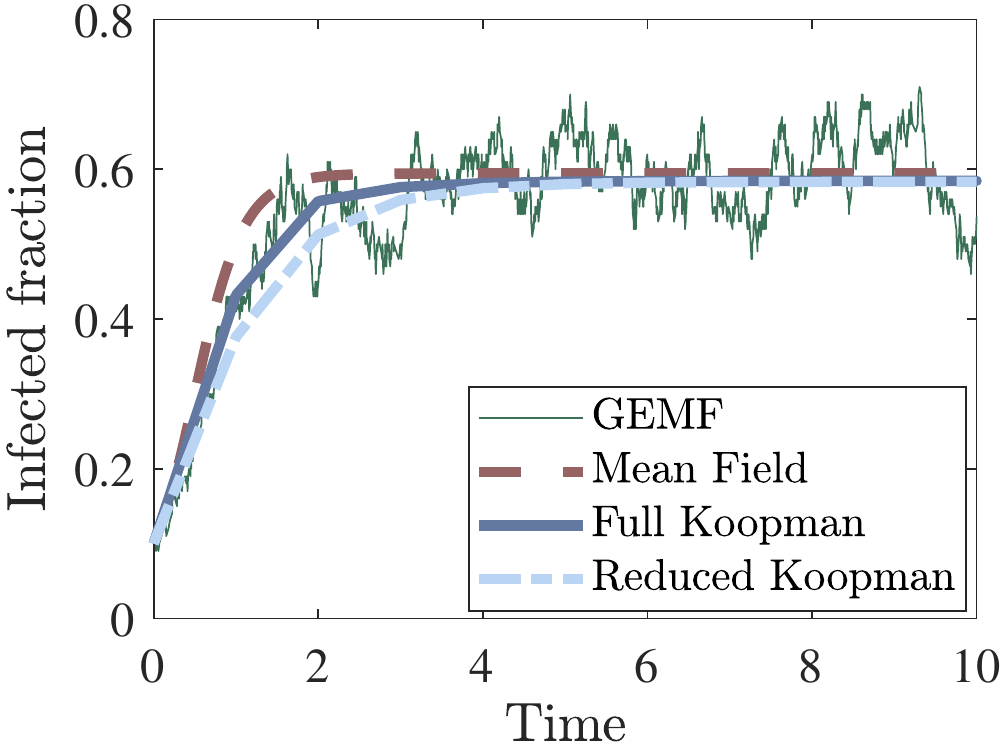}}

	\caption{Fraction of infected population for ER network.}
	\label{fig:Fraction} 
\end{figure}

\begin{figure*}
	\centering
	\subfloat{\includegraphics[clip,width=.3\columnwidth]{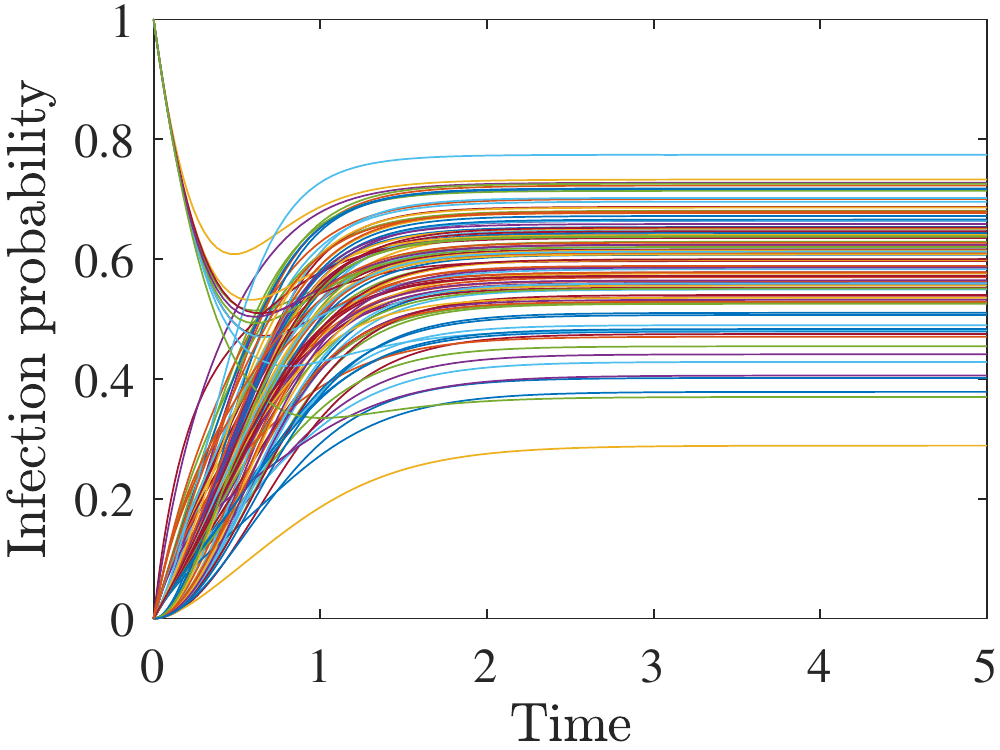}} \ \ \
	\subfloat{\includegraphics[clip,width=.3\columnwidth]{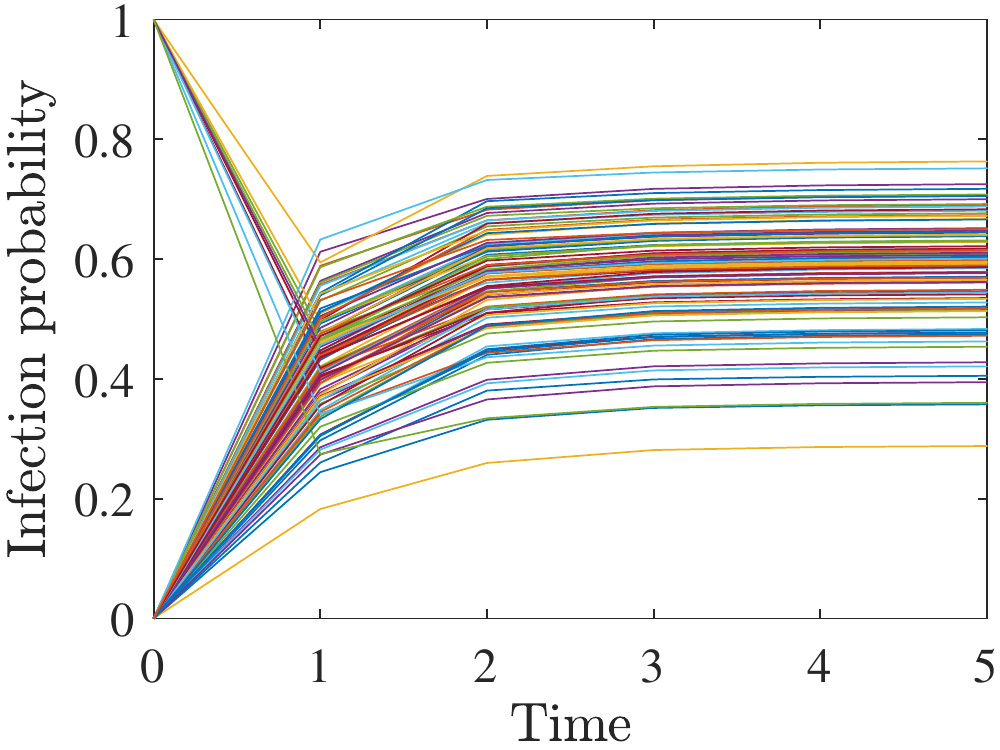}} \ \ \
	\subfloat{\includegraphics[clip,width=.3\columnwidth]{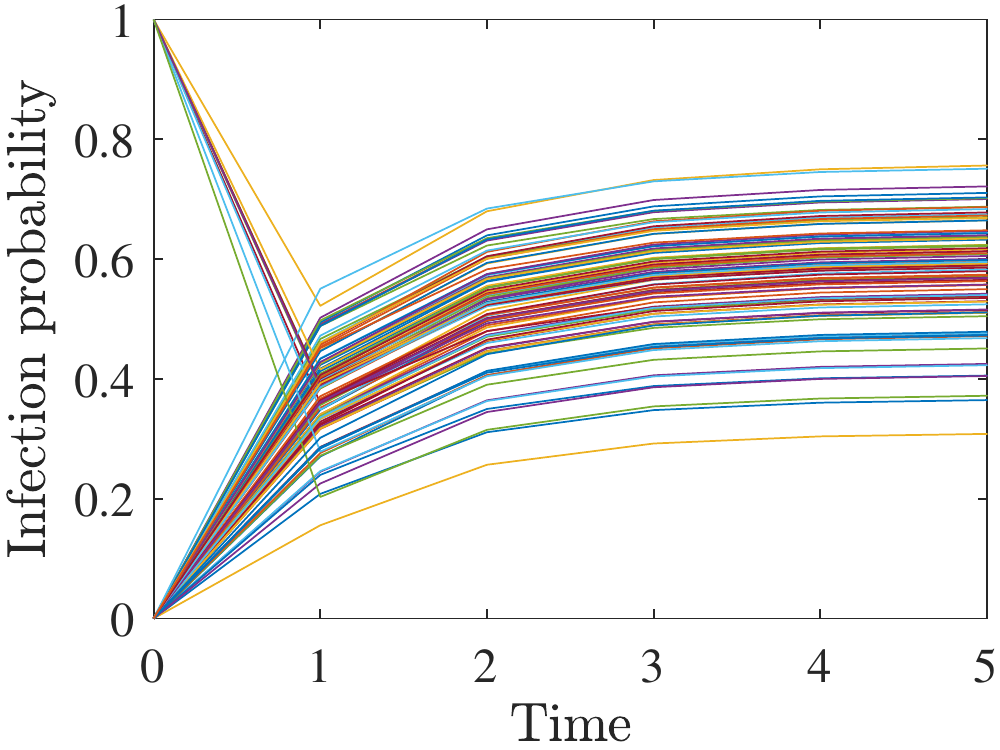}} \ \ \
		
	\caption{Probability of each individual infection corresponding to Figure \ref{fig:Fraction}. The figure shows from left the infection probabilities computed by mean field, full and reduced Koopman predictions.}
	\label{fig:IndvPrblty100} 
\end{figure*}

In this section, we consider a network of agents with the same (constant) infection rate. 
Figure \ref{fig:Fraction} shows the average fraction of infected population for $\beta_i=0.5$ with $10\%$ (randomly chosen) initial infection averaged and the predictions using mean-field model \citep{sahneh2013generalized}, full-order Koopman \eqref{eq:opt1}-\eqref{eq:opt2}, and the reduced Koopman \eqref{eq:ReducMatrix}. Koopman identification operates successfully in predicting the fraction of infected population, and the performance is comparable with the mean-field theory that is model-based and considers full information of dynamical process, system parameters, and network structure--while our approach does not. Figure \ref{fig:IndvPrblty100} illustrates the corresponding predictions for the nodal probability of infection.

We obtain the reduced order model by truncating the full order Koopman model with $r=5$ for ER and WS networks and $r=10$ for Geo network. The number of RBFs for the ER and WS networks is $200$, while it is $300$ for Geo network--we use the same values subsequently. 

\begin{figure*}
	\centering
	\subfloat[\label{fig:ErrRBFAveDeg10}]{\includegraphics[clip,width=.33\columnwidth]{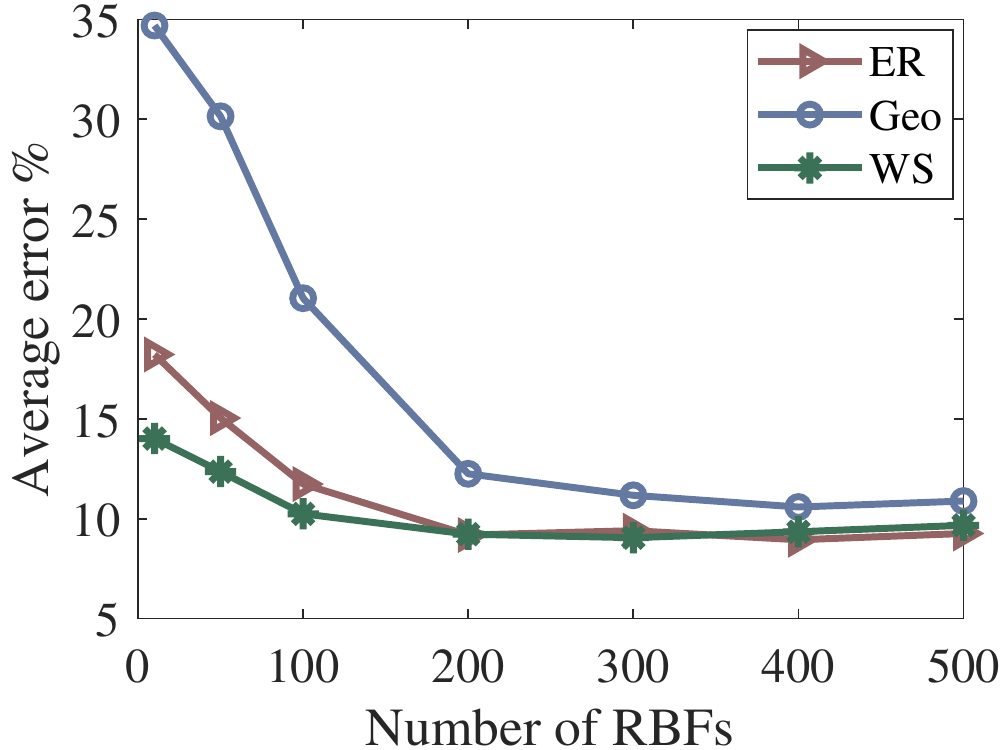}} 
	\subfloat[\label{fig:ErrReducAveDeg10}]{\includegraphics[clip,width=.33\columnwidth]{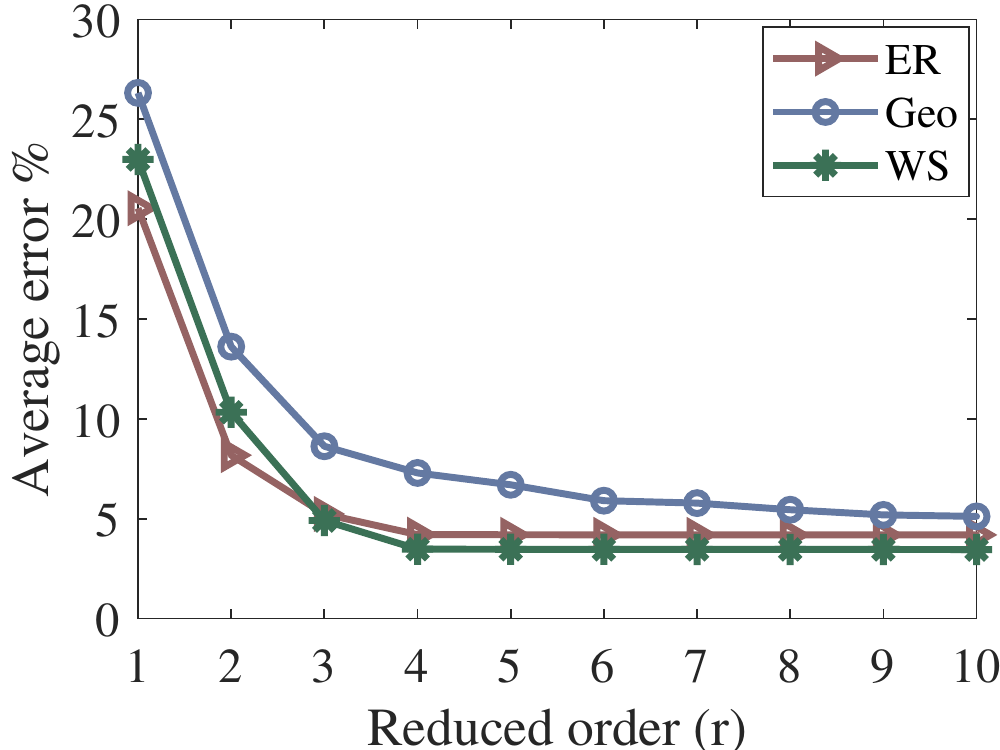}} 
	\subfloat[\label{fig:ErrRepNum}]{\includegraphics[clip,width=.33\columnwidth]{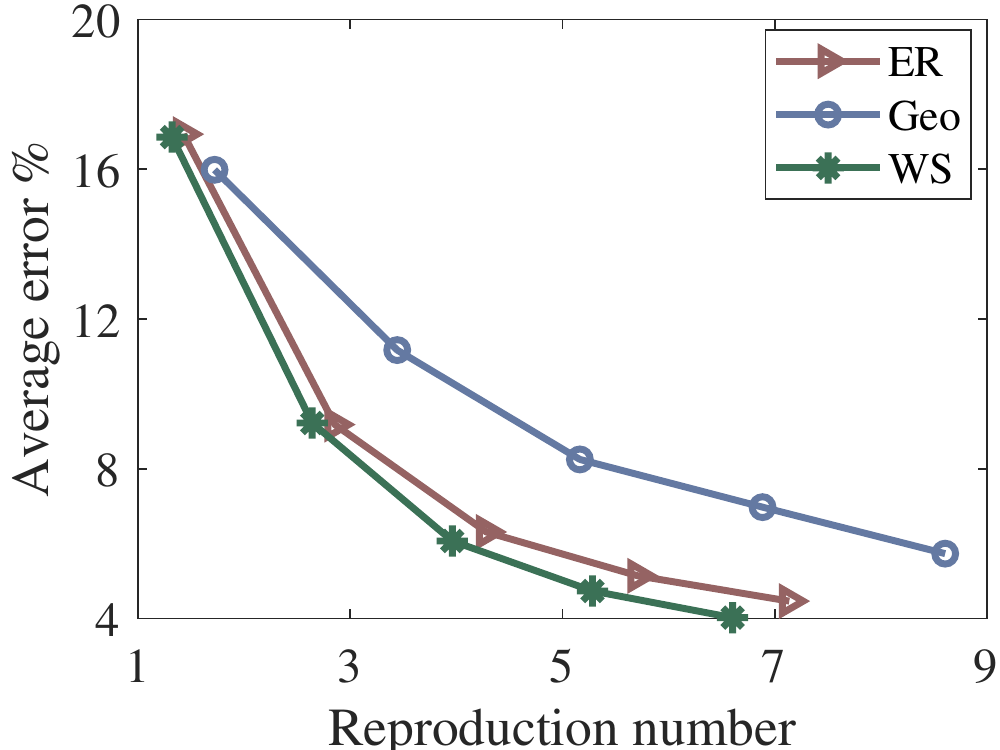}}
	\caption{Different average errors computed using 1000 randomly generated initial conditions.}
	\label{fig:ErrAveDeg10} 
\end{figure*}

We choose these numbers by investigating the prediction errors in Figure \ref{fig:ErrAveDeg10}. Each point represents the Koopman prediction error over a $t=T$ and 1000 initial conditions, averaged among the prediction errors for all agents. Hence, each error is obtained by computing two averages: one among all agents and one among all initial conditions. Firstly, we observe that the average prediction error for each of ER and WS networks remains almost unchanged by increasing the number of RBFs beyond 200; this number is more considerable for Geo networks. We stop increasing the number of RBFs beyond these values to avoid the increase of complexity and thus overfitting. Second, the evaluation of prediction error for reduced Koopman models in Figure \ref{fig:ErrReducAveDeg10} illustrates that increasing the number of Koopman modes $r$ beyond 5 for ER and WS networks and 10 for Geo network has a negligible effect on error reduction. Consequently, while Koopman embeds the stochastic nonlinear system into a high-dimensional linear model, e.g., SIS model over a network of 100 agents may be embedded into a dimension of 200 or 300, its mode decomposition can yield a much smaller, but effective, representation. The considered networks with 100 agents are successfully represented by linear models of 5 and 10 states (fifth and tenth order linear models). This implies exploring the low-dimensional manifold that describes the underlying dynamics is a promising approach for challenges of optimization and control over networks. 

We further examine average errors for different reproduction numbers obtained for different corresponding infection rates in Figure \ref{fig:ErrRepNum}. We observe that the average nodal error reduces with increasing the reproduction number $\mathcal R$. For large reproduction numbers, connections and interactions between agents grow stronger and the overall network operates more uniformly. This uniformity makes the network more predictable. Figure \ref{fig:ErrAveDeg10} also signifies that the prediction in ER and WS networks is more effortless than Geo networks; thus, we can represent them by lower-order models-- we attribute this to slower mixing dynamics and larger diameter in spatial graphs. 
 

\subsubsection{Varying input}
\begin{figure}
	\centering
	\includegraphics[clip,width=.3\columnwidth]{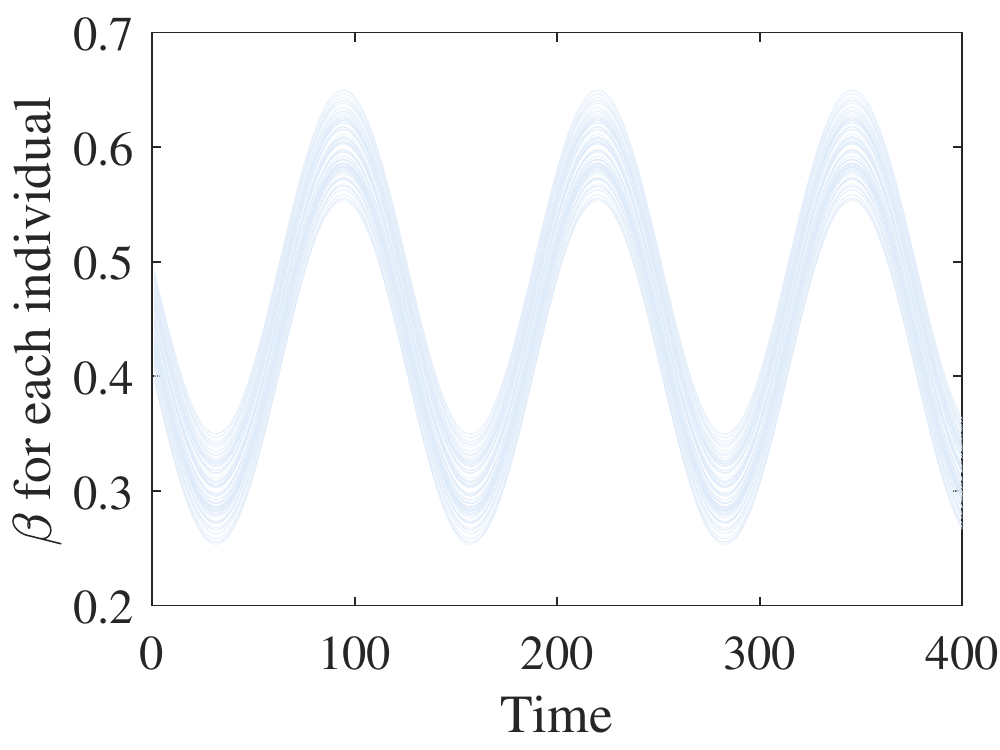}
	\caption{Time varying heterogeneous infection rate input.} 
	\label{fig:betaVariation}
\end{figure}

\begin{figure*}
\centering
\subfloat{\includegraphics[clip,width=.3\columnwidth]{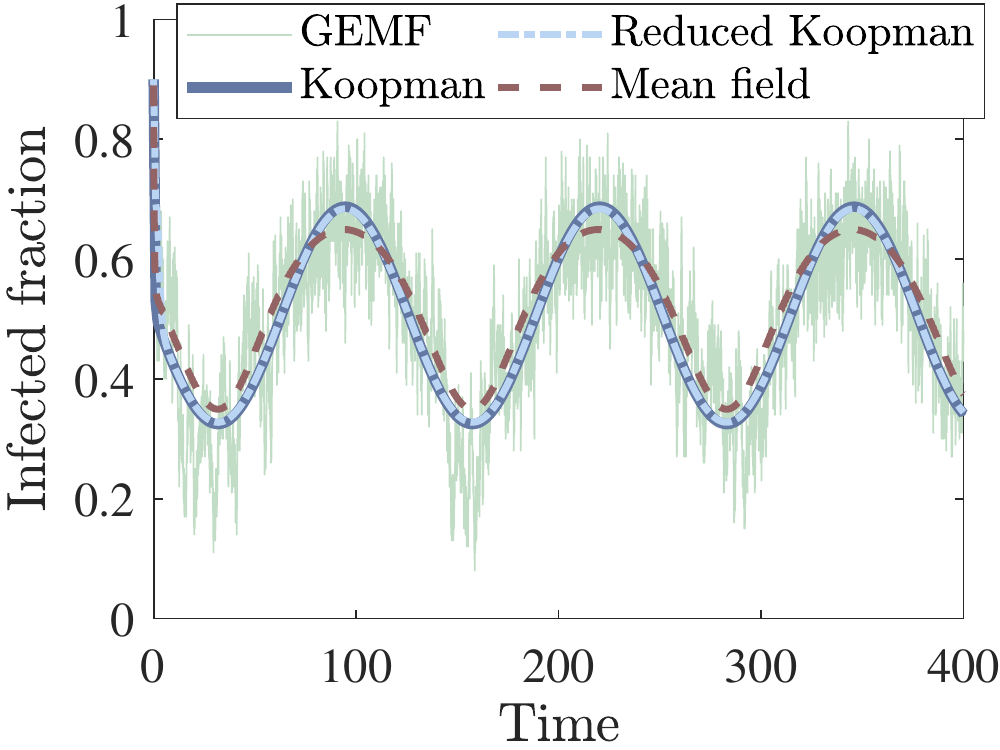}}  \ \ \ 
\subfloat{\includegraphics[clip,width=.3\columnwidth]{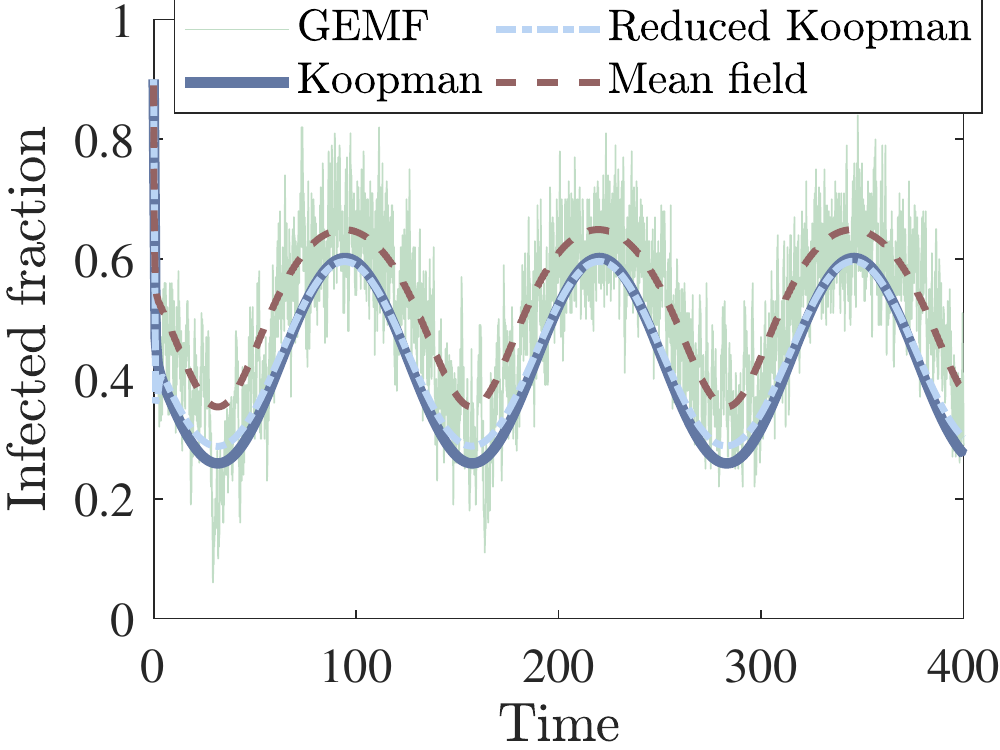}}
\caption{Prediction over ER network for varying input trained for $0.2\leq\beta\leq0.7$ (left) and $0\leq\beta\leq1$ (right).}
\label{fig:FractionVarIn} 
\end{figure*}

We examine the efficacy of the proposed approach when the infection rate varies, i.e., $\beta=\beta_0-\Delta\beta$ where $\beta_0=[{\beta_0}_i]=1$ is constant for all agents, and $u=\Delta\beta\in\mathbb R^n$ is the (heterogeneous) input vector to the system. As an example of conditions under which the infection rate is both time-varying and heterogeneous, we examine the Koopman model prediction performance in response to the oscillatory time-varying input  shown in Figure \ref{fig:betaVariation}. We train the Koopman model for two different input ranges $0.2\leq\beta\leq0.7$ and $0\leq\beta\leq1$ and compare the results. 
Figure \ref{fig:FractionVarIn} illustrates that the prediction error increases by widening the training range, highlighting the importance of input training range in the identified model accuracy.

\begin{figure*}
	\centering
	\subfloat{\includegraphics[clip,width=.35\columnwidth]{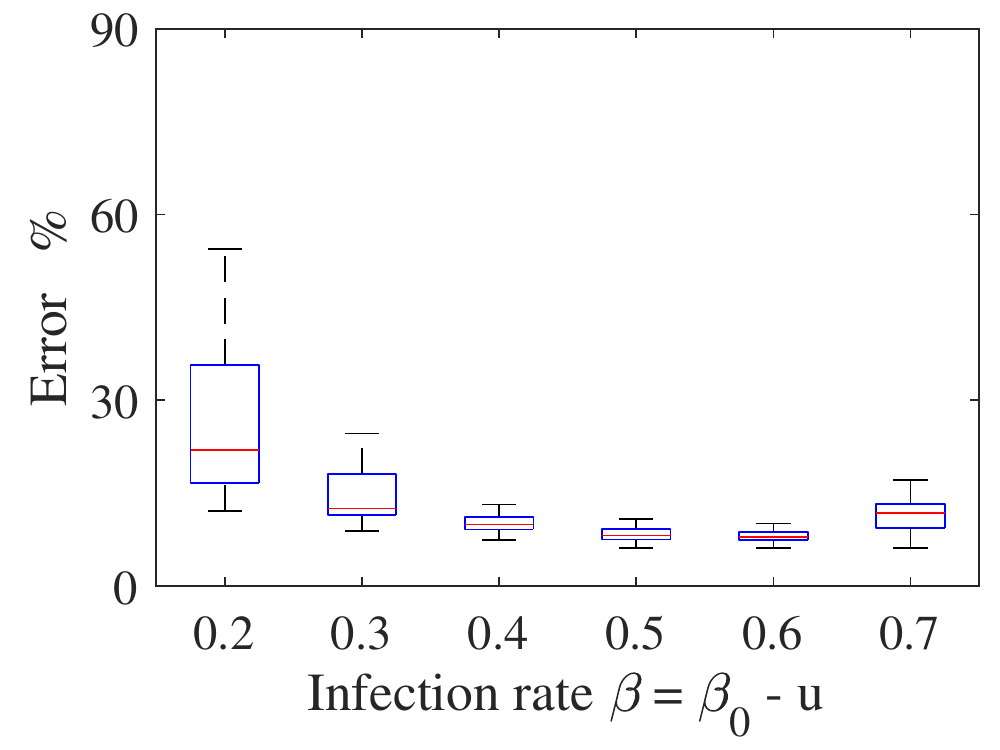}}  \ \ \
	\subfloat{\includegraphics[clip,width=.35\columnwidth]{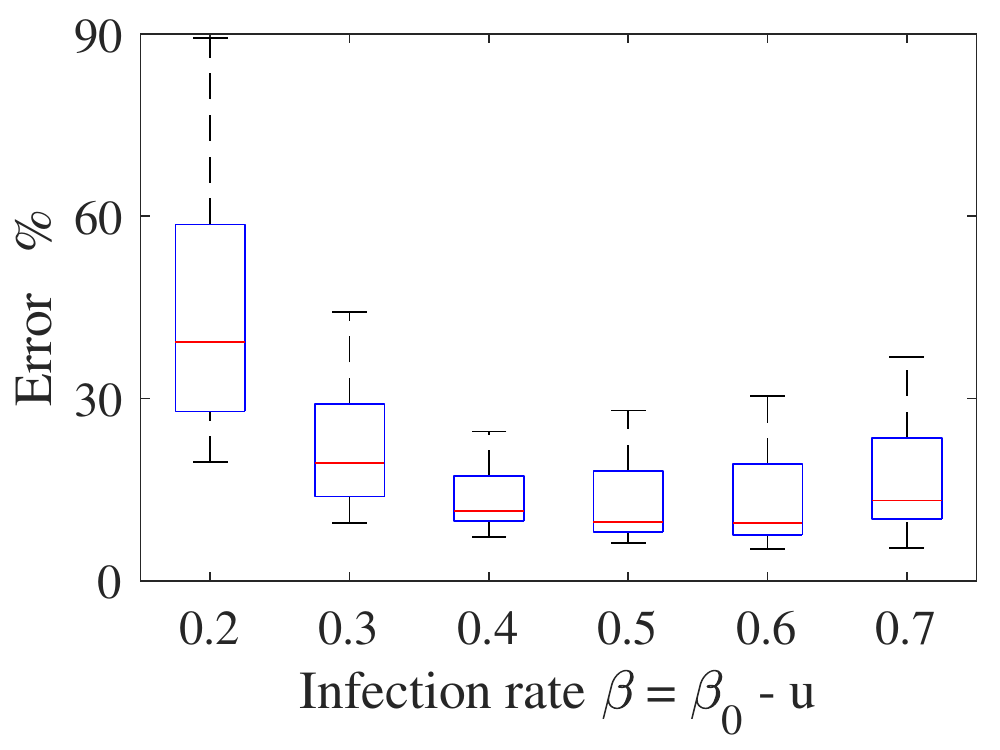}}  \\
	\subfloat{\includegraphics[clip,width=.35\columnwidth]{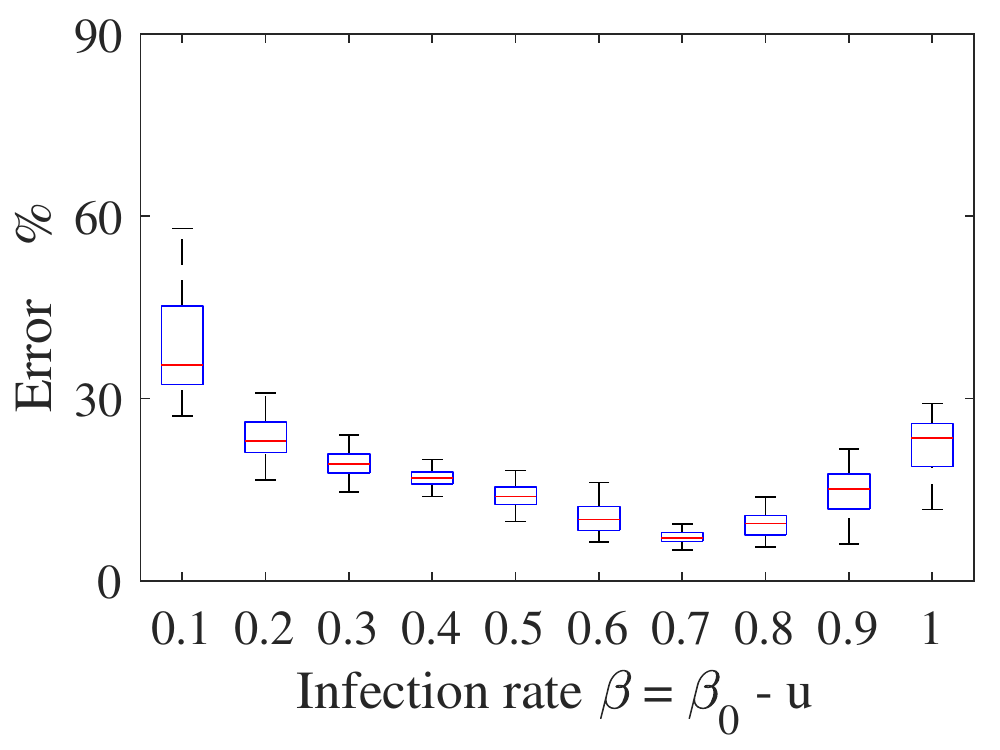}}  \ \ \
	\subfloat{\includegraphics[clip,width=.35\columnwidth]{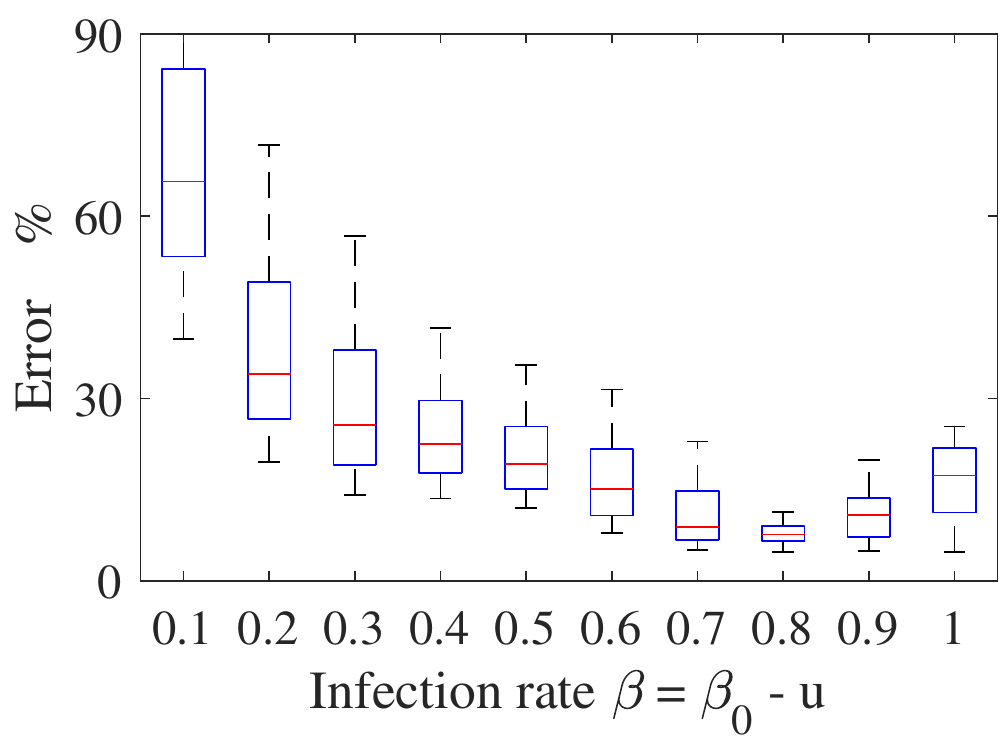}}
	\caption{Average prediction error computed by 1000 randomly generated initial conditions for different homogeneous constant inputs $u$ in the ER network. The first and second rows show the result for the training ranges $0.2\leq\beta\leq0.7$ and $0\leq\beta\leq1$, respectively, by the left column representing the full Koopman and right the reduced Koopman.  }
	\label{fig:ErrorInput} 
\end{figure*}

To further quantify our results, we probe two types of errors as metrics of performance. First, we compute the relative error when we apply a constant homogeneous input in the model trained for a given input range. The corresponding results are shown in Figure \ref{fig:ErrorInput} showing the average prediction errors for the trained Koopman models after a time period $t=T$. Figure \ref{fig:ErrorInput} indicates that the average relative error increases by approaching the boundaries of the training range. Comparing the errors corresponding to ranges $0.2\leq\beta\leq0.7$ and $0\leq\beta\leq1$ reveals that narrower input training range, i.e. a Koopman model trained in the range $0.2\leq\beta\leq0.7$, generally produces less error, thus more accurate model (see Figure \ref{fig:FractionVarIn}). Next, while the average error for most inputs in Figure \ref{fig:ErrorInput} is larger when using reduced Koopman, we observe an exception for values of $u$ corresponding to $\beta$ near 1 when the model is trained for the broader range $0\leq\beta\leq1$. This improvement is a result of balanced truncation of dynamics in the reduced Koopman model and less overfitting compared to the full Koopman model \citep{rowley2017model}. 
Second, we consider the average error for heterogeneous inputs shown in Table \ref{tabl:ErrorInput}; in this case, the error is averaged among 1000 trajectories corresponding to 1000 randomly generated initial conditions and control input vectors. Although the prediction error increases by more broader training range or further reducing the Koopman model, the full Koopman model may still experience overfitting (Figure \ref{fig:ErrorInput}). Then, the reduced Koopman's proper mode decomposition results in more accurate prediction by refining and filtering the identified model's noisy part, hence preventing overfitting.

\begin{table}[h!]
	\begin{center}
		\caption{Average error for different input training ranges}
		\label{tabl:ErrorInput}
		\begin{tabular}{c|c|c|c|c|c} 
			\multicolumn{2}{c|}{} & \multicolumn{4}{c}{Average error \%}\\
			\hline
			\multicolumn{2}{c|}{Training range} & \multicolumn{2}{c|}{$0.2\leq\beta\leq0.7$} & \multicolumn{2}{c}{$0\leq\beta\leq1$}   \\ 
			\hline
			\multicolumn{2}{c|}{Koopman type} & Full & Reduced & Full & Reduced \\
			\hline
			\multirow{3}{*}{Network} & ER & 11.55 & 24.83 & 26.48 & 65.40 \\
			& Geo & 12.50 & 28.65 & 25.73 & 59.65 \\
			& WS & 11.39 & 24.72 & 26.04 & 65.80 \\
		\end{tabular}
	\end{center}
\end{table}

\subsection{Koopman MPC for networked SIS}
\subsubsection{Limited budget problem}
In this section, we consider a linear cost function as $l_i(\mathbb E[\bar x_i])=\boldsymbol 1^T\mathbb E[\bar x_i]$ in \eqref{eq:MPC}, where $\boldsymbol1$ is the all ones vector, thus minimizing the fraction of the infected population. 
Instead of explicitly minimizing the control expenditure, we limit the total control action at each time step by a budget $u_T$ by enforcing the constraint $\boldsymbol1^Tu\leq u_T$. Furthermore, we assume the control input at each node is limited as $0\leq u_i\leq {\beta_0}_i$, so that the infection rate $\beta_i$ of each node can be neither negative nor increased beyond the initial value ${\beta_0}_i$. We impose no state constraints, i.e., $C_i(\mathbb E[\bar x_i])=0$. The problem becomes an optimal assignment of resources to mitigate the epidemic with a prediction horizon $p=3$. 

\begin{figure}
	\centering
	\subfloat{\includegraphics[clip,width=.33\columnwidth]{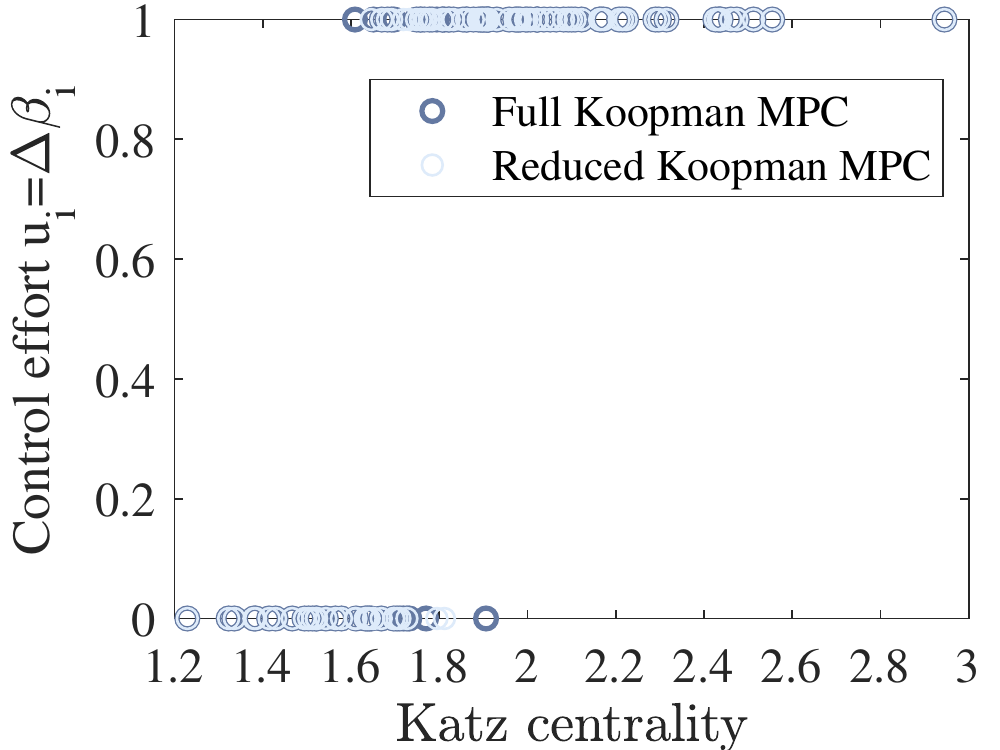}} \ \ \
	\subfloat{\includegraphics[clip,width=.33\columnwidth]{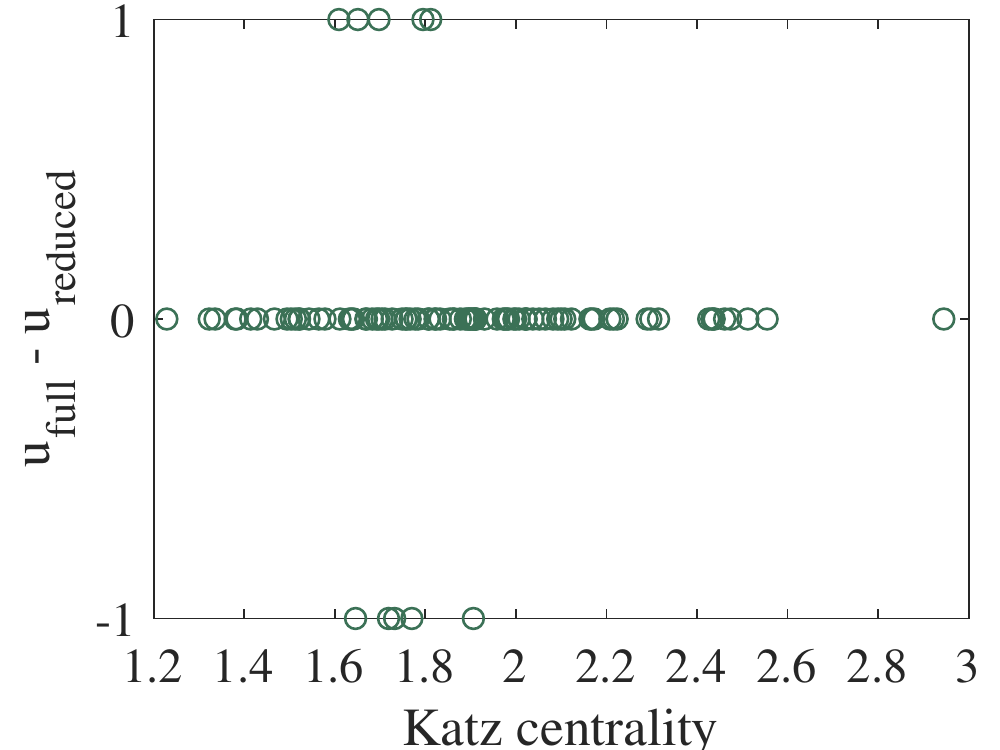}}
	\caption{Control distribution in Figure \ref{fig:MPC}. The left figure shows the control distribution, and the right shows the corresponding control input differences between full and reduced Koopman MPC. }
	\label{fig:CntrlDstrbtn} 
\end{figure}

We assume 90 percent of the population is initially infected. For comparison, we present the results of another scenario where the total available budget $u_T$ is distributed uniformly among all agents. For simulation, we set ${\beta_0}_i=1$ and $u_T=0.7\sum_{i=1}^{n}{\beta_0}_i=70$. Figure \ref{fig:MPC} illustrates a typical system response where on the one hand, a uniform resource allocation fails to mitigate epidemic by driving the system into an endemic state, and on the other hand, MPC via Koopman approaches operates successfully to halt the epidemic throughout the network. 
Both full and reduced Koopman models perform almost equally, with a slight advantage with full Koopman MPC, indicating the reduced Koopman MPC is nearly as effective as the full Koopman MPC, though being of significantly lower order. Figure \ref{fig:CntrlDstrbtn} shows the control distributions and the nodes' Katz centrality. 
The optimal control strategy in this limited budget case, with linear MPC cost function, is constant over time and distributes the total budget to nodes with the most centrality measures. Thus, the most central nodes are assigned maximum control action while the others with lesser centrality measures are left without action ($u_i=0$). 

This strategy is significant for practical use, e.g., if the control action is to vaccinate the agents, the resource allocation policy recommends vaccinating only the most central agents. We emphasize that control architecture's assessment of the resource allocation strategy and identifying the importance of nodes is accomplished exclusively by nonlinear mode decomposition of the available data, without any knowledge of system parameters or network geometry.  
Table \ref{tabl:AverageTransition} compares the average new cases of infection after applying control, obtained by averaging among trajectories of 1000 randomly selected initial conditions. We observe fewer infection cases using the full Koopman model for ER and Geo networks in the limited budget problem. However, in the WS network, the reduced-order Koopman induces fewer infection cases in the limited budget problem; an improvement by proper mode decomposition in the reduced model that reduces overfitting. Table \ref{tabl:AverageTransition} confirms controlling the epidemic in Geo network is more complicated than in ER and WS.

\begin{figure}
	\centering
	\subfloat{\includegraphics[clip,width=.33\columnwidth]{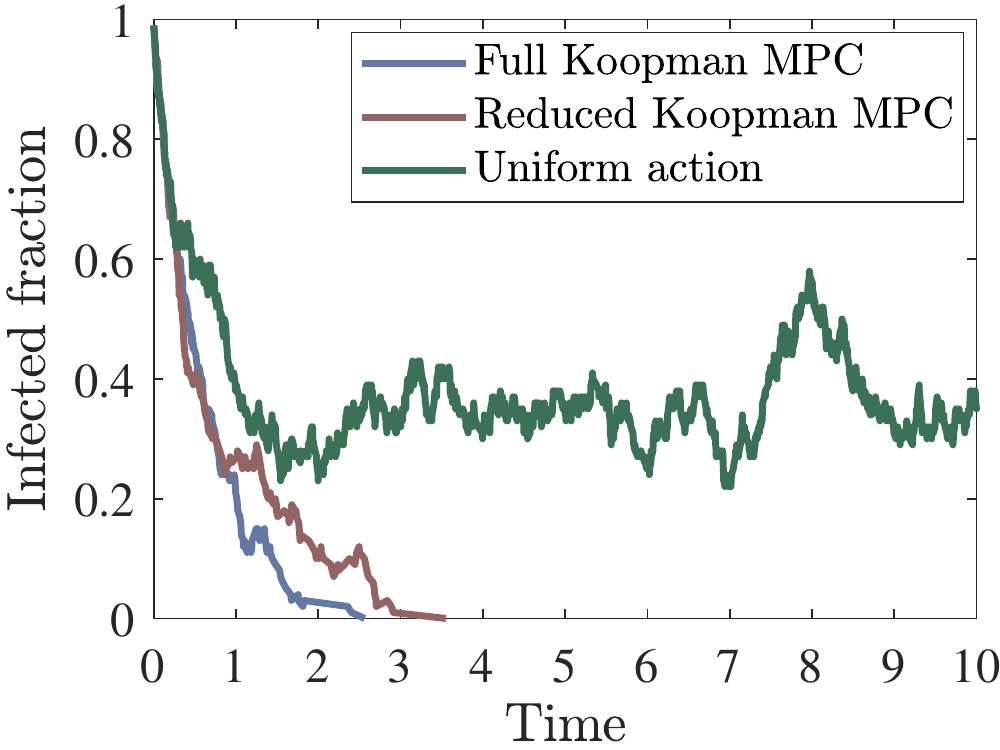}}
	\caption{Fraction of infected population under MPC with limited total budget in ER network. }
	\label{fig:MPC} 
\end{figure}
\begin{table}[h!]
	\begin{center}
		\caption{Average number of transitions S$\rightarrow$I, i.e. number of infections, after applying MPC in a network with $100$ nodes}
		\label{tabl:AverageTransition}
		\begin{tabular}{c|c|c|c|c|c} 
			\multicolumn{2}{c|}{} & \multicolumn{4}{c}{Average transition} \\
			\hline
			\multicolumn{2}{c|}{MPC strategy} & \multicolumn{2}{c|}{Limited budget} & \multicolumn{2}{c}{Minimum cost}  \\
			\hline
			\multicolumn{2}{c|}{Koopman type} & Full & Reduced & Full & Reduced  \\
			\hline
			\multirow{3}{*}{Network} & ER & 29.40 & 33.55 & 24.39 & 35.10 \\
			& Geo & 92.60 & 126.05 & 30.51 & 86.21  \\
			& WS & 64.27 & 48.11 & 33.80 & 45.13 \\
		\end{tabular}
	\end{center}
\end{table}

\subsubsection{Minimum cost problem}
In the previous subsection, the control action was concluded to be constant with time for the linear cost function. To reach a time varying resource allocation strategy, we consider a quadratic cost function as $l_i(\mathbb E[\bar x_i])=\mathbb E[\bar x_i]^T\hat Q\mathbb E[\bar x_i]+\hat q^T\mathbb E[\bar x_i]$ in \eqref{eq:MPC}, where $\hat q\in\mathbb R^n$, and $\hat Q\in\mathbb R^{n\times n}$ is positive semidefinite. Although we consider no constraint directly imposed on the total available budget, the control action of each node is still limited as $0\leq u_i\leq {\beta_0}_i=1$, and it also contributes to cost function by choosing nonzero values for $r_i$ and $R_i$ in \eqref{eq:MPC}. There is no constraint on system state too, $C_i(\mathbb E[\bar x_i])=0$. Consequently, our aim is to mitigate an existing epidemic while minimizing the costs. 

\begin{figure}
	\centering
	\subfloat{\includegraphics[clip,width=.33\columnwidth]{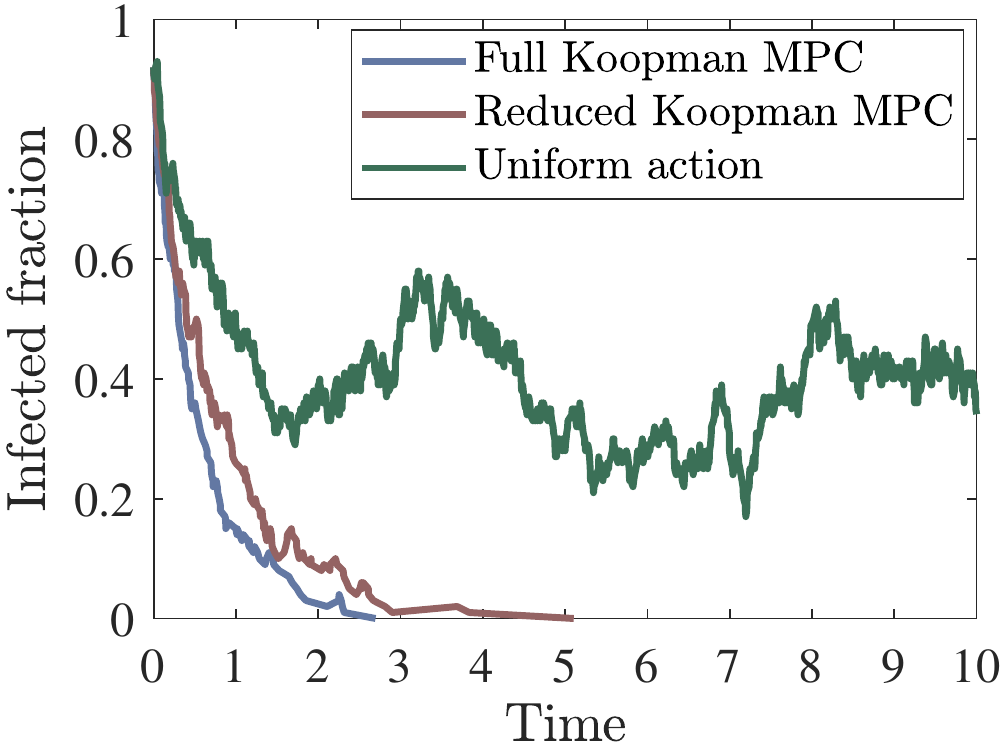}}
	\caption{Fraction of infected population under MPC with minimum cost in ER network. }
	\label{fig:MPC_MinCost} 
\end{figure}

\begin{figure}
	\centering
	\subfloat{\includegraphics[clip,width=.33\columnwidth]{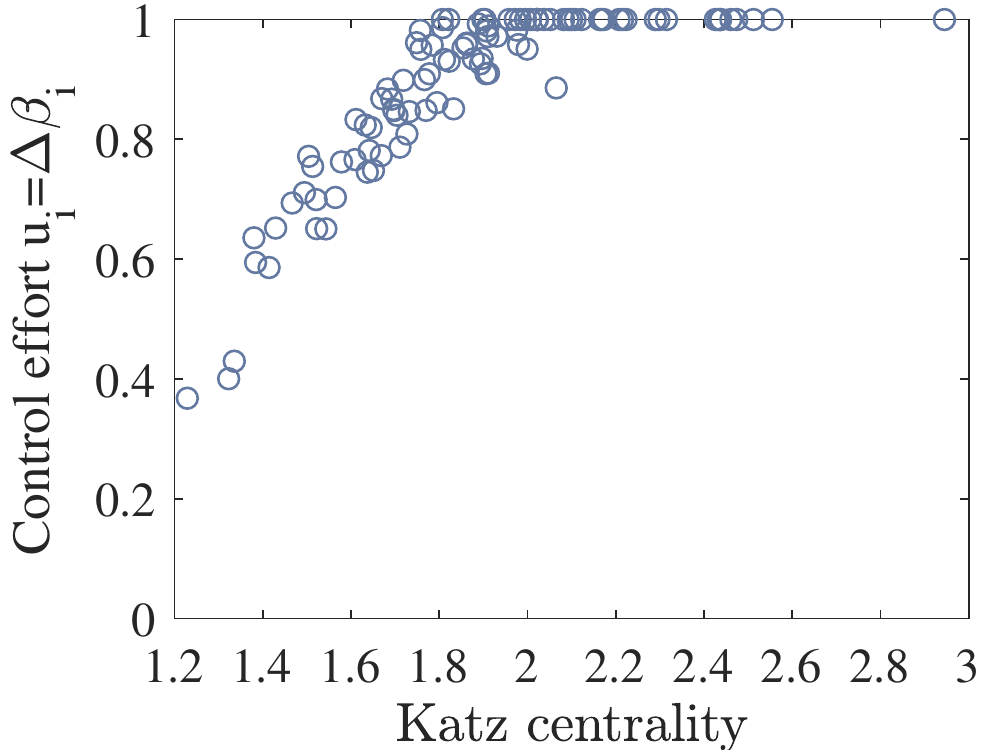}} \ \ \
	\subfloat{\includegraphics[clip,width=.33\columnwidth]{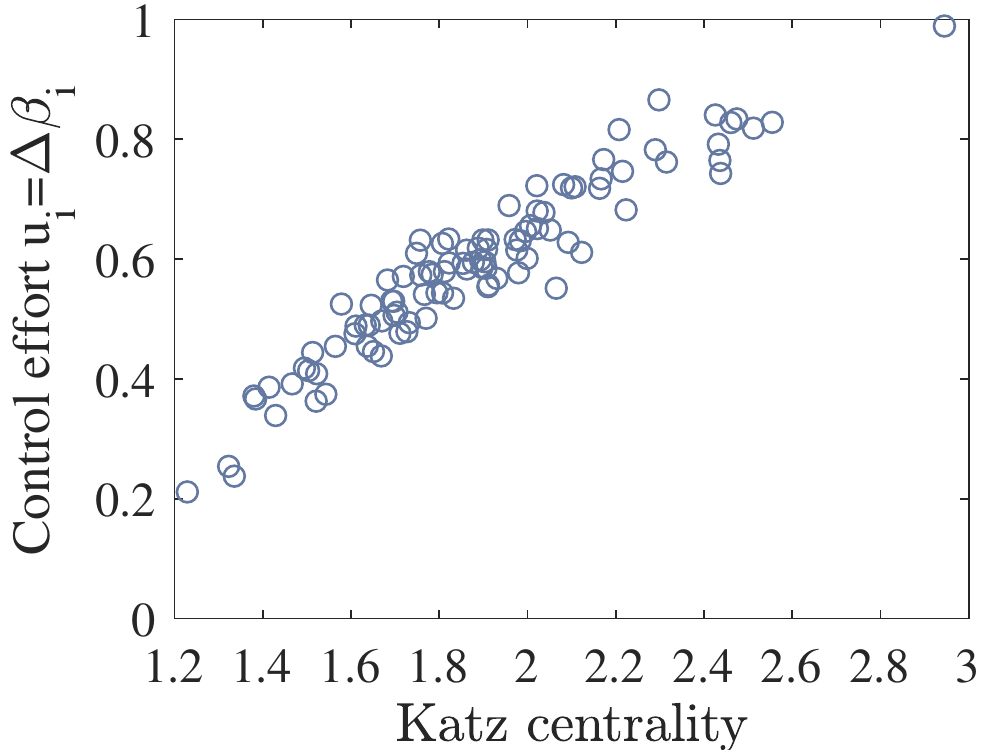}}
	\caption{Control distribution by full Koopman MPC in Figure \ref{fig:MPC_MinCost} at $t=1$ (left) and $t=10$ (right). }
	\label{fig:CntrlDstrbtnMinCost} 
\end{figure}

For numerical values we consider $\hat Q=I_{n\times n}$, $\hat q=0.5\boldsymbol 1_{n}$, $R_i=0.3I_{n\times n}$, and $r_i=0.1\boldsymbol 1_{n}$, where $I_{n\times n}$ denotes identity matrix of size $n$, and $\boldsymbol 1_n$ the all ones vector of size $n$. Figure \ref{fig:MPC_MinCost} shows a typical system response where we observe Koopman models' success in mitigating the epidemic, something that is not possible with uniform resource distribution. Moreover, while the full Koopman model performs slightly better, the reduced Koopman model performance is comparable. Figure \ref{fig:CntrlDstrbtnMinCost} indicates the control allocation of the full Koopman model for times $t=1$ and $t=10$. The reduced Koopman model decides qualitatively similar control actions (we avoid repeating similar results in the paper). 

\begin{figure}
	\centering
	\subfloat{\includegraphics[clip,width=.33\columnwidth]{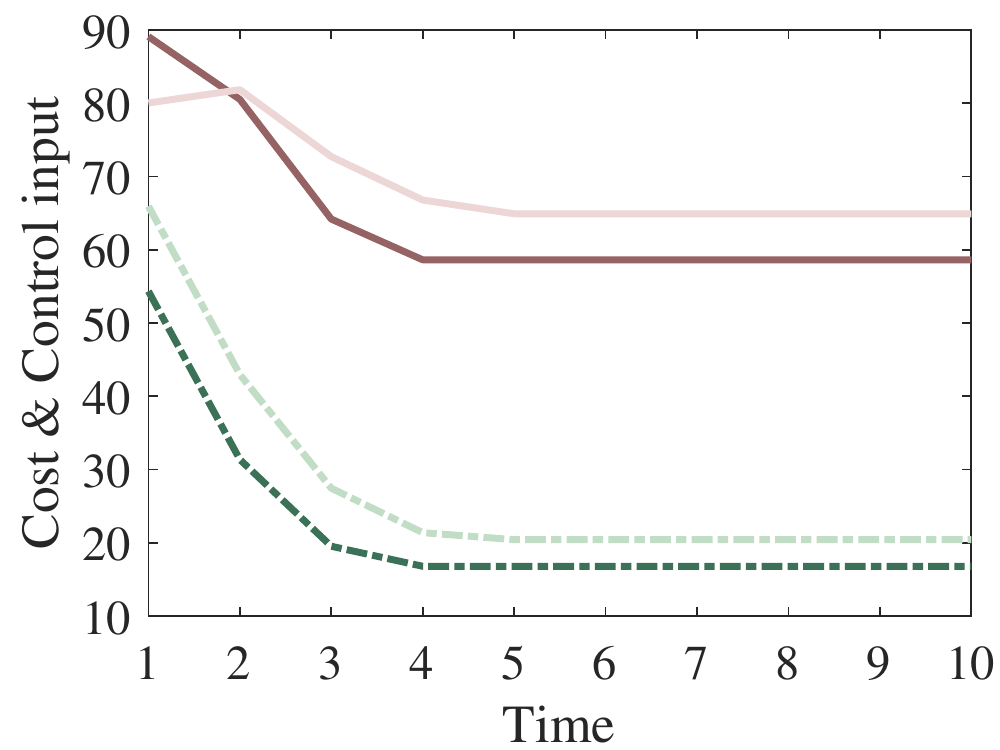}}
	\caption{Total control applied (solid lines) and cost function value (dash-dotted lines) in Figure \ref{fig:MPC_MinCost}. Dark colors show the result of full Koopman and the pale show reduced Koopman.}
	\label{fig:CostCntrl} 
\end{figure}

Figure \ref{fig:CntrlDstrbtnMinCost} illustrates that the MPC effort initially concentrates mainly on reducing the epidemic by increasing and saturating the control actions near the maximum value 1. Hence, only some nodes of small centrality measures are not assigned their maximum possible control (see Figure \ref{fig:CntrlDstrbtnMinCost} on left). With time passing and the epidemic decaying, the MPC strategy turns to give more priority to minimum control action corresponding to less budget, so that applied control inputs decrease significantly (see Figure \ref{fig:CntrlDstrbtnMinCost} on right). 
Figure \ref{fig:CostCntrl} further illustrates this by referring to time variations of the total control action, where we also plotted the MPC cost function values during the epidemics. For total control action, Figure \ref{fig:CostCntrl} verifies a nonincreasing pattern where the reduced Koopman often induces more control effort than the full Koopman except for the beginning, i.e., $t=1$. Figure \ref{fig:CostCntrl} also shows the minimum cost function value by full Koopman is smaller than that of the reduced order. Finally, we observe for the minimum cost problem in Table \ref{tabl:AverageTransition} that, after applying MPC, the full Koopman model results in fewer new infection cases than reduced one does. 



\section{Conclusion  and discussion}\label{sec:conclusion}
Modern data-driven techniques yield promising tools to identify, optimize, and control of dynamical processes over complex networks. In this work, we use operator-theoretic methods to characterize stochastic nonlinear dynamics and represent them into low-dimensional linear forms. This is beneficial to accurately predict complex networked processes through interpretable models that can be effectively utilized to reformulate the existing optimization and control problems on networks. This approach
converts the original network MPC, a nonlinear optimization problem, into a convex problem with fewer decision variables. As a specific application of the proposed method, we concluded its power to predict and control epidemic spread over networks.
Among different random graphs studied, the random geometric networks (Geo) showed more complicated features for identification and control. That is, the Geo network needs more effort compared to ER and WS networks. This is attributed to slower mixing dynamics and larger diameter in spatial graphs.

Optimization of network dynamics has a long-standing history due to its paramount importance in areas as diverse as engineering, physics, biology, the social sciences, computer science, and economics. However, this vast literature still fails to achieve a comprehensive solution for challenging features originating from nonlinear phenomena, stochastic processes, large system scale, and complex network structures. The control inputs differ from strategies adopted in \citep{Preciado2009Spectral, Mieghem2011Rmoval} that considered removing nodes and/or removing links that lead to combinatorial NP-hard problems, and similar to \citep{preciado2014optimal}, by distributing resources that promote corrective behaviors in terms of continuous properties of nodes. Moreover, instead of off-line strategies in \citet{preciado2014optimal, shakeri2015optimal, Nowzari2017Optimal, Watkins2018Optimal}, our approach is an online control strategy that monitors the system state. Therefore it provides feedback and thus possesses robustness properties against system uncertainties and exogenous disturbances, all when no knowledge of the network structure or parameters is provided.

While optimal control strategies are recently employed to solve various online control problems over networks \citep{Khanafer2014Optimal, Eshgi2016Malware, Kandhway2016Information, Mieghem2019Optimal, Dashtbali2020Game, Watkins2020MPC}, they fall short in practice. Specifically, they are based on unrealistically simplified deterministic models, have a computational burden that is intractable for large networks, and require complete knowledge of network geometry and dynamical parameters. Our proposed approach leverages the advantages of operator-theoretic methods \citep{Klus2018} to treat an original problem within a framework where the fundamental theories and practices are well developed. Furthermore, we utilize modern data-driven techniques to identify such operators for network dynamics. The success of our proposed strategy lies in the topological conjugacy \citep{LAN2013} that allows us to exploit the linearity of Koopman dynamics and tame the original nonlinear dynamics. Unlike local linearization approaches \citep{Khanafer2014Optimal}, that are valid within a (small) neighborhood of invariant sets, Koopman eigenfunctions extend the validity of the linear model into the whole basin of attraction. Furthermore we offer computationally tractable solutions, in contrary to recent works that use nonlinear models for more accurate and stable control \citep{Mieghem2019Optimal, Watkins2020MPC} with recalcitrant nonlinear programmings with requirements about the exact knowledge of underlying dynamics, model parameters, and network geometry. Hence, the importance of this work remains in establishing an approach that does not ask for often-unknown network information over and enables practical linear control strategies that are valid over the state space.

Model reductions in networks often are based on graph clustering and aggregation \citep{Cheng2021OrderReduction} with assumptions on network structures. However, network intricacies and interconnections give rise to dynamics that evolve on low-order manifolds, and operator-theoretic techniques can capture these manifolds \citep{Klus2018} efficiently. The approximation of Koopman operator using EDMD with balanced truncation represents the nonlinear dynamics of low-order manifolds by considering the most effective Koopman eigenfunctions. We use such low-order linear models to offer a tractable framework for significant control problems, such as MPC, over large networks.


In what follows, we acknowledge and discuss the limitations and possible extensions of our approach. Although EDMD is a simple approach, it only approximates the Koopman operator if the observables library is chosen appropriately. Practices such as deep learning techniques are proposed to improve this choice to better asses invariant Koopman subspaces \citep{Li2017Learning, Lusch2018DL, Otto2019Learning, Noe2020DL, Pan2020PhysInfr}. Therefore future inclusions of these techniques may result in more accurate prediction and control of network processes. Moreover, we assume no prior knowledge of the system dynamics, but when possible, physics-informed machine learning techniques \citep{Karniadakis2021, Pan2020PhysInfr} can reduce data volume and reach better accuracy, faster training, and improved generalization.
On the other hand, if we have information on the network geometry, we can utilize the sparse reduced-order modeling approach to full-state estimation \citep{Loiseau2018Sensor} by only monitoring the states of a few numbers of agents. This will yield a more practical version of this work, since we are not always provided with full measurement of the network state. Another extension of this work can be made by multi-scale identification of underlying dynamics by collecting data of agent groups instead of individual agents. Although by the group-based strategy we only estimate the state in each group state, not each agent, it is effective particularly over large networks by significantly reducing the computational burden \citep{Darabi2021GroupGEM}.

\bibliographystyle{chicago}
\bibliography{refs}
\end{document}